\documentclass[]{emulateapj-rtx4}

\usepackage{natbib,color} 
\bibpunct{(}{)}{;}{a}{}{,}

\DeclareRobustCommand{\ion}[2]{%
\relax\ifmmode
\ifx\testbx\f@series
{\mathbf{#1\,\mathsc{#2}}}\else
{\mathrm{#1\,\mathsc{#2}}}\fi
\else\textup{#1\,{\mdseries\textsc{#2}}}%
\fi}

\shorttitle{Temporal Evolution of the Inverse Evershed Flow}
\shortauthors{Beck, C.; Choudhary, D.P.}

\begin{document}
\title{Temporal Evolution of the Inverse Evershed Flow} 


\author{C. Beck}
\affil{National Solar Observatory (NSO), Boulder, USA}
\author{D.P. Choudhary}
\affil{Department of Physics \& Astronomy, California State University, Northridge, USA}




\begin{abstract}
The inverse Evershed flow (IEF) is an inflow of material into the penumbra of sunspots in the solar chromosphere that occurs along dark, elongated superpenumbral fibrils extending from about the outer edge of the moat cell to the sunspot. The IEF channels exhibit brightenings in the penumbra, where the supersonic IEF descends to the photosphere causing shock fronts with localized heating. We used an 1-hr time-series of spectroscopic observations of the chromospheric spectral lines of \ion{Ca}{ii} IR at 854\,nm and H$\alpha$ at 656\,nm taken with the Interferometric BIdimensional Spectrometer at the Dunn Solar Telescope to investigate the temporal evolution of IEF channels. Complementary information on the photospheric magnetic field was obtained from observations with the Facility InfraRed Spectropolarimeter at 1083\,nm and the Helioseismic and Magnetic Imager. We find that individual IEF channels are long-lived (10--60\,min) and only show minor changes in position and flow speed during their life time. Initiation and termination of IEF channels takes several minutes. The IEF channels with line-of-sight velocities of about 10\,km\,s$^{-1}$ show no lasting impact from transient or oscillatory phenomena with maximal velocity amplitudes of only about 1\,km\,s$^{-1}$ that run along them. We could not detect any clear correlation of the location and evolution of IEF channels to local magnetic field properties in the photosphere in the penumbra or moving magnetic features in the sunspot moat. Our results support a picture of the IEF as a field-aligned siphon flow along arched loops. From our data we cannot determine if their evolution is controlled by events at the outer end in the moat or at the inner end in the penumbra.
\end{abstract}

\section{Introduction}

The three-dimensional topology of the magnetic field of sunspots supports mass motions and wave propagation along field lines at all heights from the photosphere to the corona \citep{staude1999,solanki2003,khomenko+collados2015}. The dominant mass motions are the Evershed flow \citep[EF;][]{evershed1909} in the photosphere that is directed outward from the sunspot center and the inverse Evershed flow \citep[IEF;][]{evershed1909a,maltby1975,boerner+kneer1992} at chromospheric heights in the opposite direction that sets in at a height of about 500 km. On rare occasions, photospheric counter-Evershed flows were also observed that were associated with high chromospheric activity \citep{louis+etal2014a,siutapia+etal2017}. 

Most wave motions observed in both photospheric and chromospheric lines were found to originate in the umbra or at the umbral boundary and to propagate outwards \citep{zirin+stein1972,antia+etal1978,rouppevandervoort+etal2003}. Besides these systematic motions, numerous transient phenomena were detected such as chromospheric jets and anomalous flows in sunspots associated with local dynamical magnetic phenomena \citep{katsukawa+etal2007,morton2012,louis+etal2014b}. In higher atmospheric layers, coronal rain and flocculent flows were observed which are believed to result from rapid cooling of material at those heights that subsequently slides down along the magnetic field lines \citep{vissers+etal2012, ahn+etal2014}.


\begin{figure*}
\hspace*{.6cm}\resizebox{17.2cm}{!}{\includegraphics{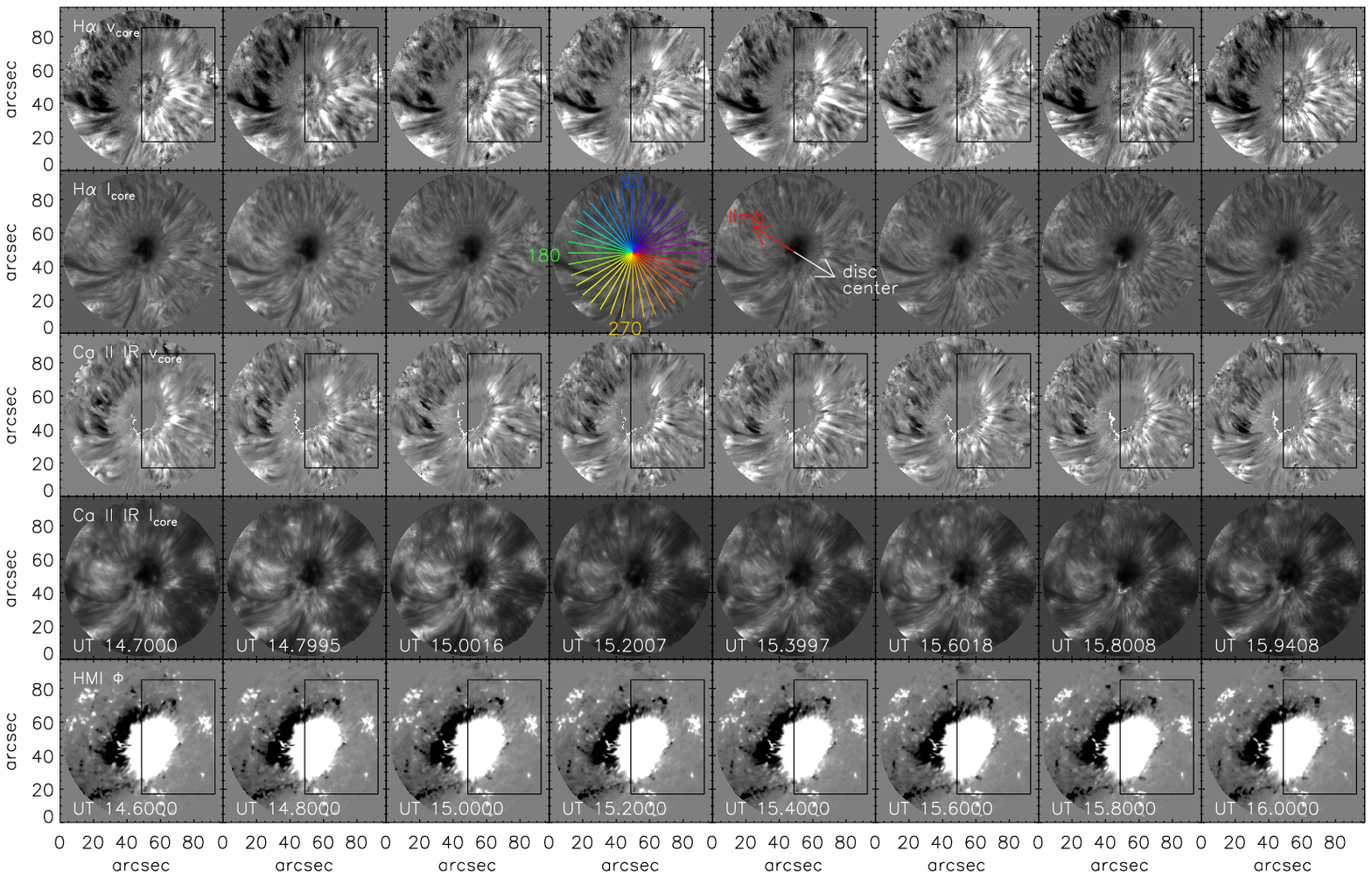}}\\$ $\\$ $\\$ $\\
\centerline{\resizebox{12cm}{!}{\includegraphics{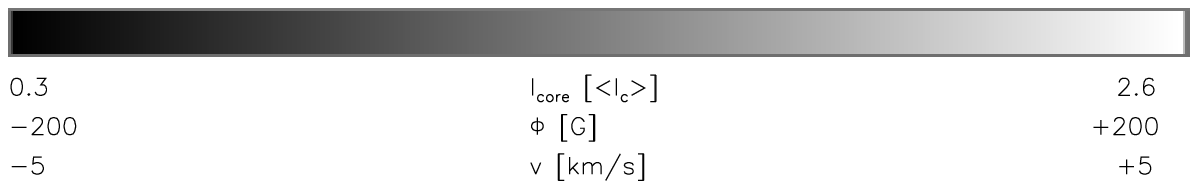}}}
\caption{Aligned HMI and IBIS data. Bottom to top: HMI line-of-sight flux $\Phi$ in the range of $\pm 200$\,G, IBIS line-core intensity and line-core velocity ($\pm 5$\,km\,s$^{-1}$) of \ion{Ca}{ii} IR, IBIS line-core intensity and line-core velocity ($\pm 5$\,km\,s$^{-1}$) of H$\alpha$. Time increases from left to right as indicated in the bottom two rows. The black rectangles mark the center side region with the clearest IEF channels. The colored lines in the 4th row and column show the location of the radial cuts around the sunspot center with a 10$^\circ$ spacing, where 0$^\circ$ corresponds to right and 90$^\circ$ to up. The symmetry line of the sunspot is tilted by 32$^\circ$ to the right. An animation of the time series is available in the online material. }\label{fig1}
\end{figure*}

The photospheric EF is one of the earliest detected phenomena in sunspots and has been studied extensively using both space and ground-based high-resolution instrumentation \citep{rimmele1995a, ichimoto+etal2007}. It starts at so-called penumbral grains, i.e., bright points at the umbral end of penumbral filaments, flows intermittently radially outward following elevated, nearly horizontal magnetic field lines and ends mostly at the outer penumbral boundary \citep{johannesson1993,rimmele+marino2006,beckthesis2006a,beck2008,franz+schliche2009}. The flow velocity decreases systematically with increasing height in the atmosphere already across the line formation region of photospheric lines \citep{hirzberger+kneer2001}. The driving mechanism of the EF could be either a siphon flow or magneto-convection \citep{westendorpplaza+etal1997,schlichenmaier+etal1998,scharmer+etal2011,rempel2012}, or a mixture of both. Photospheric counter-Evershed flows, i.e., inward flows in the photospheric penumbra, on the other hand were explained invoking a siphon flow mechanism by \citet{siutapia+etal2018}. The photospheric flows usually stop at about the outer penumbral boundary \citep{rezaei+etal2006}, but sometimes were seen to extend outward with supersonic velocities \citep{martinezpillet+etal2009}. The continuation of the EF beyond the penumbral boundary into the moat region has been studied using both local correlation tracking techniques of magnetic elements and spectropolarimetric observations that showed an association with moving magnetic features \citep{rimmele1995, borrero+etal2004, cabrerasolana+etal2006,vargasdominguez+etal2007}. 

At chromospheric heights, wave patterns dominate the rapid temporal evolution of the umbra and penumbra. In umbral flashes \citep[UFs;][]{beckers+tallant1969,rouppevandervoort+etal2003,felipe+etal2010,houston+etal2018}, large parts of the umbra oscillate coherently leading to shock fronts in the umbral chromosphere. In the penumbra, running penumbral waves (RPWs) are seen, which presumably are magneto-acoustic waves generated in the photosphere, that propagate radially outwards at chromospheric layers and dissipate energy at chromospheric or coronal heights \citep{tziotziou+etal2006,bloomfield+etal2007,jess+etal2013,jess+etal2015, prasad+etal2015,grant+etal2018}.

\begin{figure}
\hspace*{.8cm}\resizebox{8.cm}{!}{\includegraphics{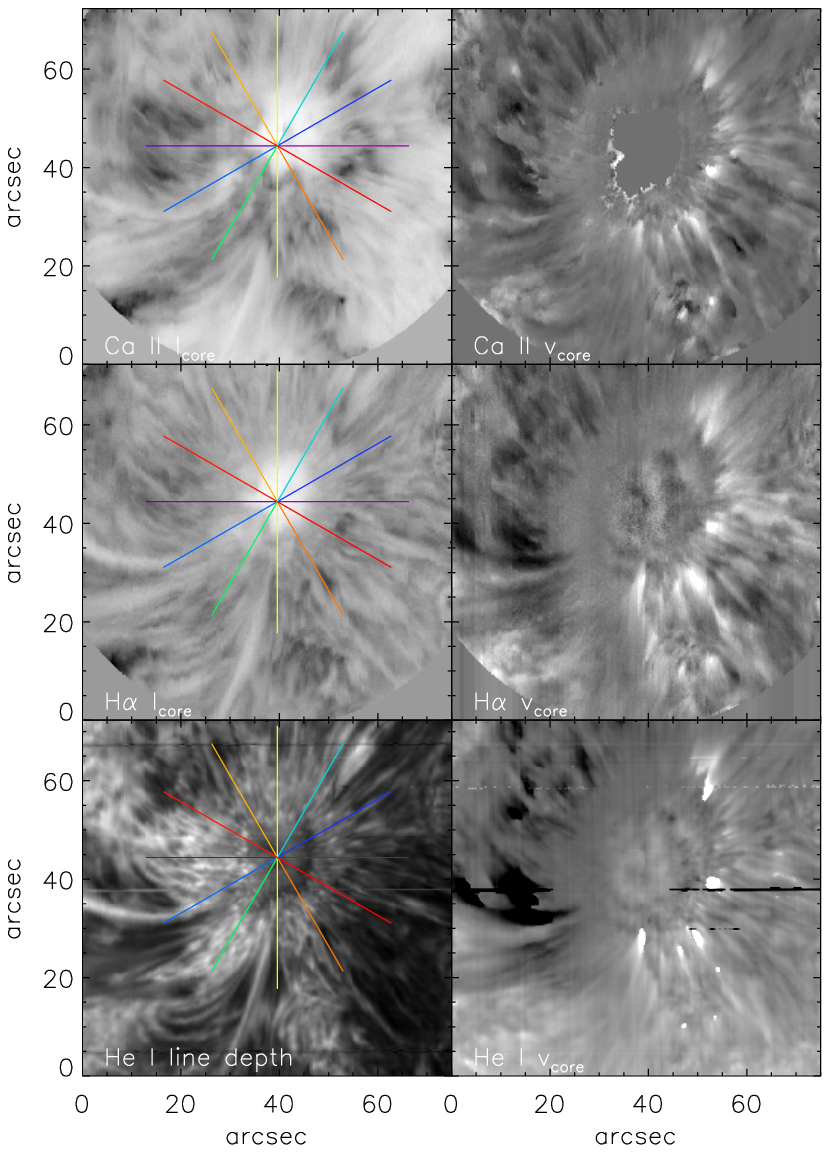}}\\$ $\\$ $\\$ $\\
\centerline{\hspace*{.8cm}\resizebox{8.cm}{!}{\includegraphics{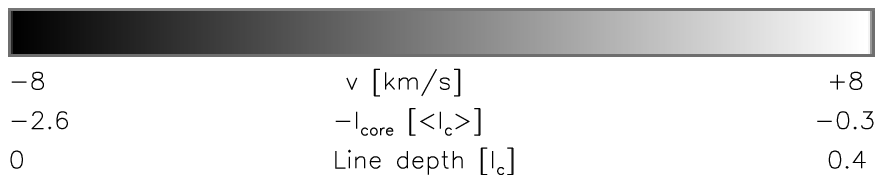}}}
\caption{FIRS data and IBIS pseudo-scan maps. Left column, bottom to top: relative line depth of \ion{He}{i} 1083\,nm, line-core intensity of H$\alpha$ and \ion{Ca}{ii} IR in reverse gray scale. Right column, bottom to top: LOS velocity of \ion{He}{i}, H$\alpha$ and \ion{Ca}{ii} IR. The colored lines in the left column indicate the radial cuts with 30$^\circ$ spacing.}\label{fig1a}
\end{figure}
\begin{figure}
\resizebox{8.8cm}{!}{\includegraphics{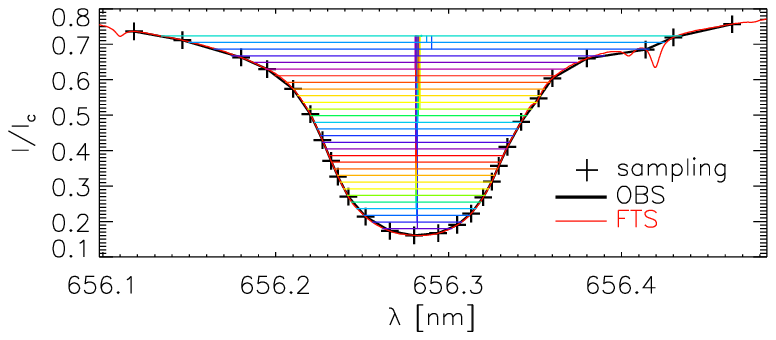}}\\
\resizebox{8.8cm}{!}{\includegraphics{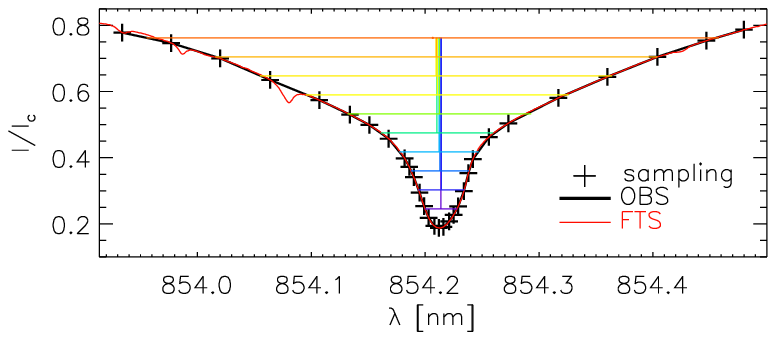}}
\caption{Example of the bisector analysis for H$\alpha$ (top panel) and \ion{Ca}{ii} IR (bottom panel) for the average profile of one scan. The horizontal colored lines indicate the bisectors that are defined by bisector intensity value, central position and width.}\label{fig1b}
\end{figure}

Steady and longer-lived mass motions are observed as the IEF at chromospheric heights and above that transport material into the sunspot from the superpenumbral boundary and beyond as first observed by \citet{evershed1909a} and later extensively studied by \citet{stjohn1913}. \citet{stjohn1913} found IEF velocities of about 3\,km\,s$^{-1}$, which were much higher than EF velocities seen at that time, a fact that was confirmed by \citet{bones+maltby1978}. Spectroscopic measurements based on H$\alpha$ spectra showed higher IEF velocities of up to 50 kms$^{-1}$ \citep{beckers1962,haugen1969, alissandrakis+etal1988, maltby1975,dere+etal1990,choudhary+beck2018}, which exceed the chromospheric sound speed. The transition region IEF velocities were found to be of the order of a few ten km\,s$^{-1}$ \citep{alissandrakis+etal1988, dere+etal1990, kjeldsethmoe+etal1993, teriaca+etal2008}. A siphon flow is thought to be the most likely mechanism among the several theoretical ideas that were considered to drive the IEF such as gravitationally driven downflows or moving flux tubes \citep{beckers1962,montesinos+thomas1997,teriaca+etal2008}. 

The properties of the flow channels carrying the IEF were derived in recent years through various techniques using modern spectropolarimetry and adaptive optics. From H$\alpha$ filtergrams it was found earlier that the the IEF channels contain compression shocks, last for about 70 minutes to more than two hours, and that in many cases old channels are replaced by new ones \citep{maltby1975, maltby1997}. The flow channels are inclined to the local vertical in the range of 30$^\circ$ to 80$^\circ$ \citep{haugen1969,beck+choudhary2019}. Comprehensive studies of the temporal evolution of both EF and IEF have been carried out using observations with the Universal Birefringent Filter at the Dunn Solar Telescope in the photospheric \ion{Fe}{i} line at 557.6 nm and the chromospheric H$\alpha$ line \citep{georgakilas+christopoulou2003}. Their results show a complex pattern of mass motions and waves at both photospheric and chromospheric heights with the EF, IEF, RPWs and an additional outward radial motion of ``velocity packets'' at a speed of 5--6 km $^{-1}$. These ``velocity packets'' were defined by \citet{georgakilas+christopoulou2003} as small patches of velocities with an opposite sign to the IEF of a few Mm extent in the outer penumbra and the moat region. A generic problem of some older studies on the IEF is that because of the spatial and temporal resolution of the data available at the time these different phenomena can get mixed up.

In this paper, we study the temporal evolution of the IEF using high-resolution spectroscopic and spectropolarimetric observations in multiple spectral lines that originate at different atmospheric heights. Our data analysis clearly shows both the IEF and wave phenomena such as RPWs and their interaction in different diagnostics. However, here we shall focus only on the temporal character of the IEF and postpone the study of the interaction with different types of waves. The semi-stationary IEF is only impacted in a minor way by the transient waves and in general shows velocities that are an order of magnitude larger than the wave amplitudes. 
\begin{figure*}
\begin{minipage}{12cm}
\resizebox{12cm}{!}{\includegraphics{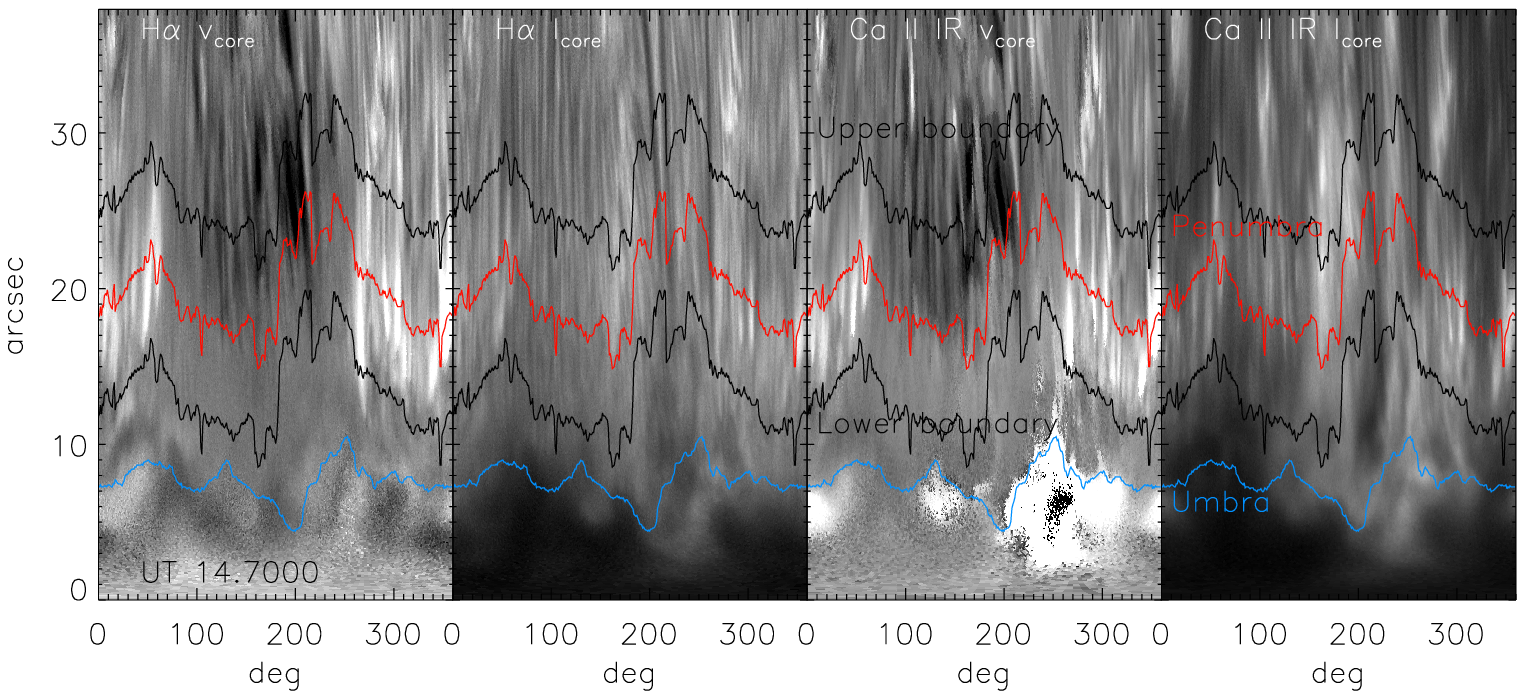}}\\$ $\\$ $\\$ $\\
\resizebox{12cm}{!}{\includegraphics{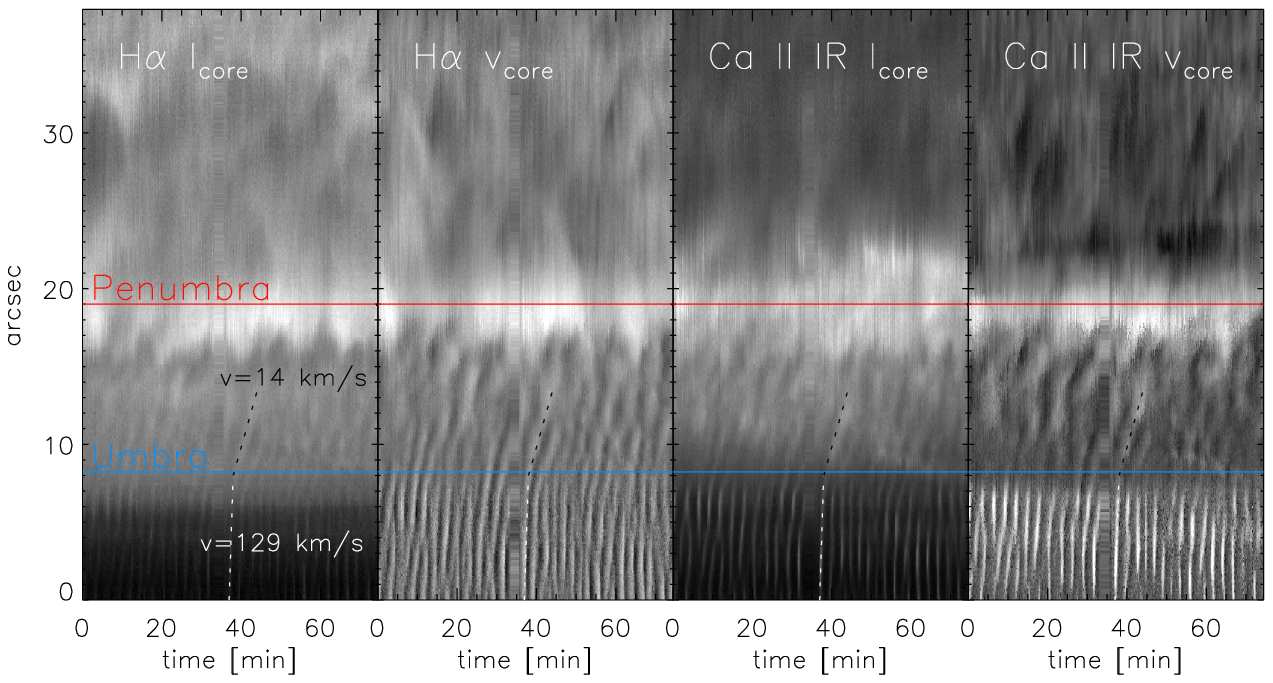}}\\$ $\\$ $\\
\end{minipage}\hspace*{.5cm}
\begin{minipage}{3cm}
\vspace*{-1cm}\resizebox{2cm}{!}{\includegraphics{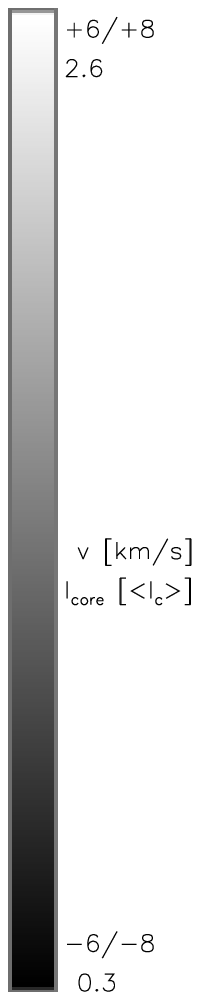}}
\end{minipage}
\caption{Line parameters along radial cuts. Bottom panel: temporal evolution at azimuth 0$^\circ$ for (left to right) line-core intensity and velocity ($\pm 8$\,km\,s$^{-1}$) of H$\alpha$, and line-core intensity and velocity ($\pm 8$\,km\,s$^{-1}$) of \ion{Ca}{ii} IR. The blue and red horizontal lines indicate the average radial distance of the outer umbral and penumbral boundary. The inclined dashed black (white) line indicates an apparent radial propagation speed of 14\,km\,s$^{-1}$ (129\,km\,s$^{-1}$) in the penumbra (umbra). Top panel: azimuthal variation of the same quantities for the first spectral scan with a $\pm 6$\,km\,s$^{-1}$ range for the velocities. The blue and red lines indicate the umbral and penumbral boundary. The black lines indicate the radial range that encloses most IEF channels. \label{fig2}}
\end{figure*}

To avoid a confusion of different phenomena, we define for the current study the chromospheric IEF channels as superpenumbral, radially oriented fibrils that connect the penumbra with the superpenumbral boundary and that exhibit a significant flow velocity (e.g., Figures \ref{fig1} and \ref{fig1a} and the animation). As in our previous paper, we define the downflow points of IEF channels as the locations close to the penumbral boundary with about maximal velocity and an enhanced brightness due to shock fronts \citep[][in the following Paper I]{choudhary+beck2018}. The RPWs are the concentric dark and bright circles or circle arcs in the intensity and velocity images with the sunspot as the center and show up as inclined ridges in the penumbra in space-time plots. We differentiate in addition between IEF channels and superpenumbral filaments by the latter permanently exhibiting a strongly reduced intensity and flows along a large part of their length. There are only a few such filaments in the field of view (FOV) going to the left and down during the period of our observations (Figures \ref{fig1} and \ref{fig1a}). The definition of the ``velocity packets'' follows the one of \citet{georgakilas+christopoulou2003} given above.

Section \ref{sec_obs} describes our data sets whose analysis is detailed in Section \ref{sec_ana}. Our results are presented and summarized in Sections \ref{sec_res} and \ref{sec_summ}, respectively. Section \ref{sec_disc} discusses our findings, while Section \ref{sec_concl} presents our conclusions.

\section{Observations}\label{sec_obs}
We observed the decaying active region (AR) NOAA 12418 on 16 September 2015 with the Interferometric BIdimensional Spectrometer \citep[IBIS;][]{cavallini2006,reardon+cavallini2008} and the Facility InfraRed Spectropolarimeter \citep[FIRS;][]{jaeggli+etal2010} at the Dunn Solar Telescope \citep[DST;][]{dunn1969, dunn+smartt1991}. The FOV covered the isolated leading sunspot of the AR located at about $x,y = -500^{\prime\prime},-340^{\prime\prime}$ at a heliocentric angle of about 43$^\circ$. The trailing opposite polarity towards the East had already decayed to diffuse plage and network.  

IBIS sequentially scanned the two chromospheric spectral lines of H$\alpha$ at 656\,nm and \ion{Ca}{ii} IR at 854.2\,nm in its spectroscopic mode with a non-equidistant spectral sampling of 27 and 30 wavelength points, respectively. The exposure time was 40\,ms with a total cadence for one spectral scan of 11.2\,s. From UT 14:42 until 15:56, 400 spectral scans were acquired. The 11 spectral scans Nos.~180 to 190 from UT 15:16 to 15:17 were impacted by clouds and were all replaced with the scan No.~179. The IBIS FOV spanned a circular aperture with a diameter of 95$^{\prime\prime}$ at a spatial sampling of 0\farcs095\,pixel$^{-1}$.

With FIRS we recorded vector spectropolarimetric data in a wavelength range from 1081.37 to 1085.28\,nm with a dispersion of 3.84\,pm\,pixel$^{-1}$ that covered the photospheric Zeeman-sensitive \ion{Si}{i} line at 1082.7\,nm and the chromospheric \ion{He}{i} line at 1083\,nm. We obtained one large-scale spatial map of 250 steps with a step width of 0\farcs3 from UT 14:42 until 14:55. The spatial sampling along the 72$^{\prime\prime}$ long slit of 0\farcs3 width was 0\farcs15\,pixel$^{-1}$. The total integration time per step was only 2\,s to focus on the thermodynamics of the \ion{He}{i} line instead of its polarization signal or any derived chromospheric magnetic field.

The ground-based data are complemented with the Milne-Eddington inversion results for the photospheric magnetic field (inclination $\gamma$, field strength $B$, LOS flux $\Phi$) derived from observations by the Helioseismic and Magnetic Imager \citep[HMI;][]{scherrer+etal2012} on-board the Solar Dynamics Observatory \citep[SDO;][]{pesnell+etal2012}. These HMI data cover the whole solar disk with a spatial sampling of about 0\farcs5 at a cadence of 12\,min. 
\begin{figure}
$ $\\$ $\\
\hspace*{.8cm}\resizebox{8.cm}{!}{\includegraphics{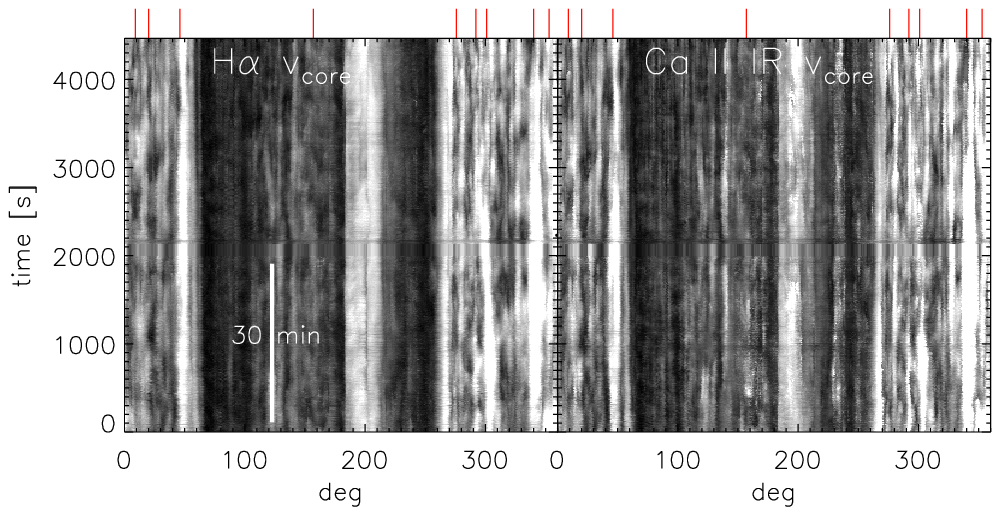}}\\$ $\\$ $\\$ $\\
\hspace*{.8cm}\resizebox{8.cm}{!}{\includegraphics{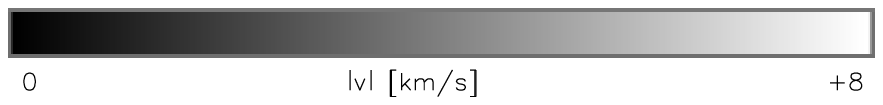}}

\caption{Temporal evolution of the unsigned maximal LOS velocities as a function of azimuth. Left/right panel: LOS velocity of H$\alpha$/\ion{Ca}{ii} IR from 0 to 8\,km\,s$^{-1}$. The maximal velocities were determined between the black lines indicated in the upper panel of Figure \ref{fig2} for each azimuth angle in each spectral scan. The vertical white bar in the left panel indicates a duration of 30\,min. The short red bars at the top indicate the azimuth angles displayed in Figure \ref{fig4}.}\label{fig3}
\end{figure}

All IBIS spectral scans were aligned to each other through a correlation of subsequent images of the continuum intensity $I_c$ in each spectral line and a subsequent shift in $x$ and $y$ with $I_c$ of H$\alpha$ as reference. The same approach was used to align the HMI data. The FOV of the ground-based data was already almost stationary because of the real-time correction by the adaptive optics system of the DST \citep{rimmele2004}.

Figure \ref{fig1} shows aligned IBIS and HMI data at the 12\,min cadence of the latter.
\begin{figure}
\resizebox{8.8cm}{!}{\hspace*{.5cm}\includegraphics{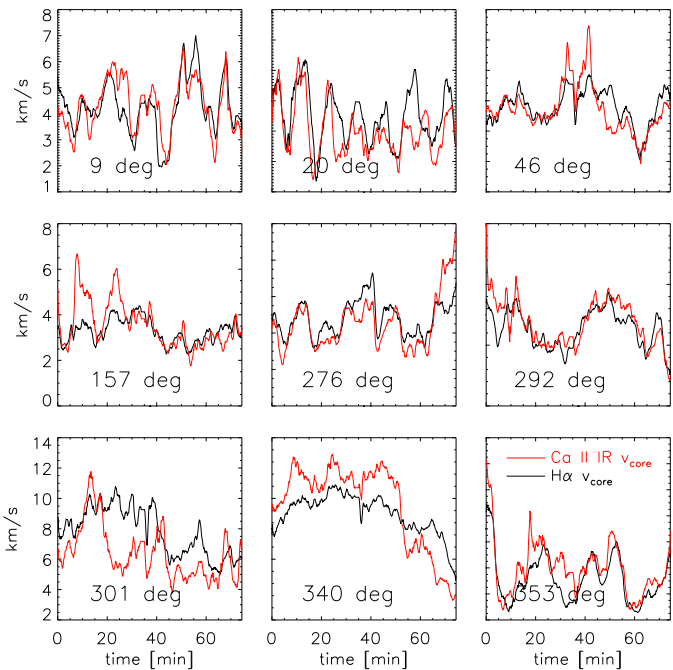}}\\$ $\\$ $\\
\caption{Temporal evolution of the absolute velocity for nine azimuth angles. Black/red lines: LOS velocity of H$\alpha$/\ion{Ca}{ii} IR. The azimuth angle is indicated in each panel.}\label{fig4}
\end{figure}

\section{Data Analysis}\label{sec_ana}
\subsection{FIRS data at 1083\,nm}
We inverted the polarimetric spectra of the \ion{Si}{i} line at 1082.7\,nm with the Stokes Inversion based on Response functions code \citep[SIR;][]{cobo+toroiniesta1992}. We used a single magnetic component with the magnetic field properties field strength $B$, inclination $\gamma$ and azimuth $\phi$ and the LOS velocity constant with optical depth for all pixels with a significant polarization signal above about 1\,\% of $I_c$. Temperature was allowed to vary on two nodes and an additional contribution of unpolarized stray light was used. For all pixels without significant polarization signal, a single non-magnetic component was used instead.

For the direct comparison of the FIRS and IBIS data, we created a pseudo-scan map from the IBIS time series \citep[e.g.,][Appendix B.2]{beck+etal2007}. A simulated slit with a width of 0\farcs3 and a length of 72$^{\prime\prime}$ was stepped across the IBIS FOV in steps of 0\farcs3. At each scan step of FIRS, the data below the location of the simulated slit at that moment was cut out from the IBIS spectral scan closest in time to the FIRS scan step. The creation of the pseudo-scan map then only depends on providing the initial position of the simulated slit inside the IBIS FOV in $x$ and $y$ at the first step. Figure \ref{fig1a} shows the resulting spatially and temporally aligned FIRS and IBIS pseudo-scan maps for the line-core intensities and line-core velocities of \ion{He}{i} at 1083\,nm, H$\alpha$ and \ion{Ca}{ii} IR. 
\begin{figure*}
\resizebox{14cm}{!}{\includegraphics{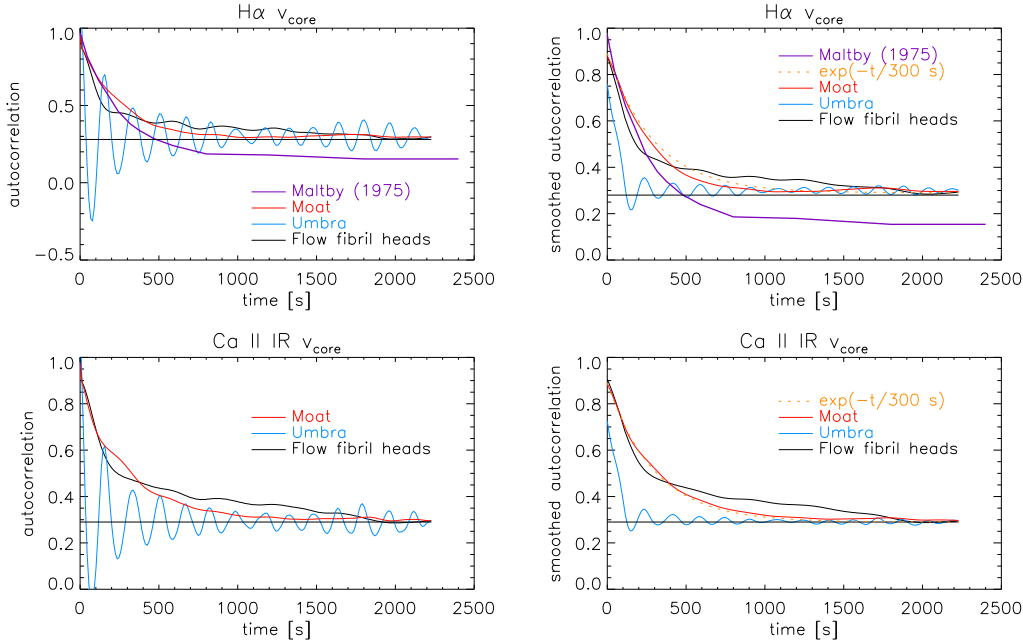}}
\caption{Auto-correlation times. Top row: temporal auto-correlation for the H$\alpha$ LOS velocity in the IEF region (black line), the umbra (blue line) and the outer moat (red line) without (left panel) and with smoothing over the dominant period of the umbral oscillations (right panel). Bottom row: the same for \ion{Ca}{ii} IR. The orange dotted lines in the right column indicate a pure exponential decay with a time constant of 300\,s. The purple line in the top row corresponds to the similar determination of temporal correlation in \citet{maltby1975}, his Figure 7. }\label{fig5}
\end{figure*}

\subsection{IBIS H$\alpha$ and \ion{Ca}{ii} IR Spectra}
We ran a bisector analysis over all H$\alpha$ and \ion{Ca}{ii} IR spectra from IBIS. The bisector analysis (see Figure \ref{fig1b}) determined the bisector velocity (central position), bisector intensity (intensity value) and bisector width (length of the bisector) for 30 (10) bisector levels in steps of 3 (10)\,\% of the relative line depth $LD  = (I_c - I_{core}) / I_c$ for H$\alpha$ (\ion{Ca}{ii} IR). The average observed spectrum of each spectral scan was first matched to the reference spectrum from the Fourier Transform Spectrometer atlas \citep[FTS][]{kurucz+etal1984} with a correction for the transmission of the IBIS pre-filter as described for instance in \citet{beck+etal2019a}. The spectra were also re-sampled to a denser wavelength grid prior to the bisector analysis. As the range of the spectral scanning did not reach continuum levels, the first wavelength point of each spectral line was substituted for $I_c$. The line depth and the corresponding relative intensity values at fractional line depth were calculated separately for each and every spectrum, i.e., $LD = LD(x,y)$, and thus vary across the FOV and with time. We used the bisector velocities at 85 (70)\,\% line depth for H$\alpha$ (\ion{Ca}{ii} IR) in the following as ``line-core'' velocities. They are more reliable and robust than levels closer to the line core because of the flatness of the profile of H$\alpha$ around the line core and the occurrence of emission peaks in \ion{Ca}{ii} IR, especially in the umbra. We still have set the umbral velocities of \ion{Ca}{ii} IR to zero in some plots as they are often spurious even at 70\,\% line depth.

The left four panels of the animation anim\_combined\_small.mp4 show the line-core intensity and velocity of H$\alpha$ and \ion{Ca}{ii} IR over all 400 spectral scans, while the rightmost column shows the temporal and azimuthal derivative of the H$\alpha$ velocity. In addition to the IEF channels in and near the penumbra, various other features such as umbral oscillations, running penumbral waves, large-scale filaments, and a potential reconnection event can be seen. 

\subsection{Extraction of Radial Cuts}
For the analysis of the results, we extracted all relevant quantities such as spectral line parameters, bisector values and inversion results on radial cuts. These cuts were set to start at the center of the sunspot and then sample the azimuthal variation on an angular step width of 1 degree. The zero point for the angle is to the right and the angle increases counter-clockwise. The symmetry line of the sunspot with about zero LOS velocities, i.e., the direction perpendicular to the connection between sunspot and disk center, goes from about 60$^\circ$ to 240$^\circ$ in azimuth. Each cut extended over a length of 38$^{\prime\prime}$ on 600 points with a spatial radial sampling of $\approx$ 0\farcs06 per point. A subset of the cuts is indicated in the middle panel of Figure \ref{fig1} and the left column of Figure \ref{fig1a}. The same cuts were extracted from all spectral scans of IBIS, the FIRS data and the IBIS pseudo-scan maps.

We determined the radial distance of the umbral and penumbral boundary for each azimuth angle by thresholds in the continuum intensity (blue and red lines in the top panel of Figure \ref{fig2}). The average umbral and penumbral radius were 8$^{\prime\prime}$ and 19$^{\prime\prime}$, respectively. For the automatic detection of the location of IEF channels in all spectral scans we also defined a minimal and maximal radial distance inside which the majority of the IEF channels are visible (black lines in top panel of Figure \ref{fig2} and black lines in Figure \ref{fig11} below). Local LOS velocity extrema outside of these boundaries were not counted and are considered to correspond to events other than IEF channels. The IEF channels on the limb side are seen further away from the sunspot center because of the projection effects. 

\section{Results}\label{sec_res}
\subsection{Temporal evolution of IEF channels}
\subsubsection{Life time}
To estimate the life time of individual IEF channels, we used the temporal evolution for each azimuth angle as shown in the lower panel of Figure \ref{fig2} for an azimuth of 0$^\circ$. We determined the maximum value of the unsigned LOS velocity within the radial range for each azimuth angle given by the black lines in the upper panel of Figure \ref{fig2}. 

\begin{figure*}
\resizebox{14cm}{!}{\includegraphics{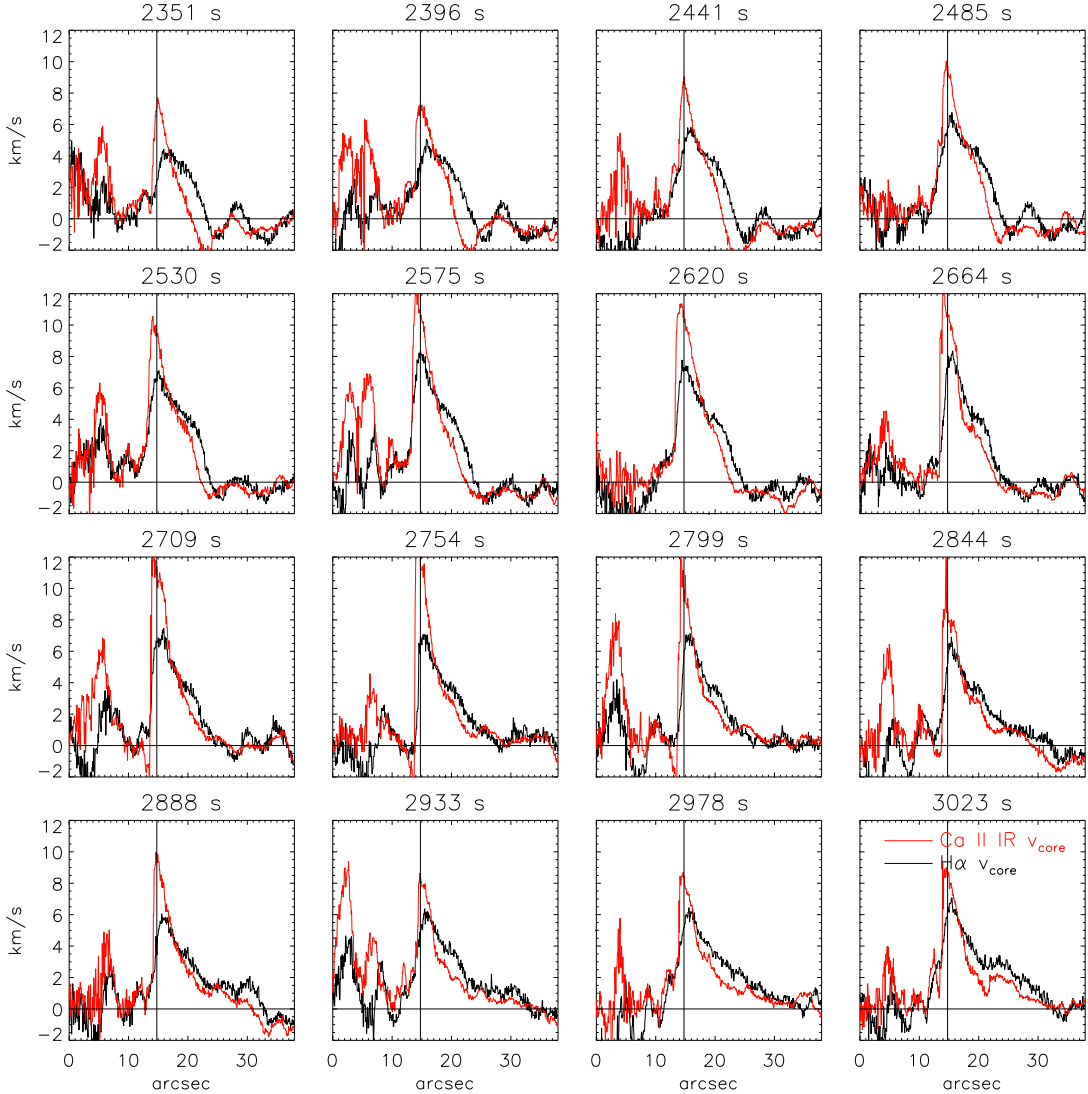}}\\$ $\\$ $\\
\caption{Temporal evolution of the LOS velocity at the start of an IEF channel at UT 15:21 at an azimuth angle of 327$^\circ$. Black/red lines: LOS velocity of H$\alpha$/\ion{Ca}{ii} IR. The black vertical lines mark the location of maximum flow speed at the first time step. Time increases from top to bottom across rows and left to right in each row.}\label{fig6}
\end{figure*}

Figure \ref{fig3} shows these maximal LOS velocities as a function of the azimuth angle and time. Continuous vertical white streaks of increased velocities indicate the extended presence of an IEF channel. Their length varies from about 10\,min to over 60\,min. Some IEF channels exist at about the same place for the whole duration of the time series. The majority of the channels seems to exist for significantly more than 10\,min. There are several occasions where an existing IEF channel is replaced by a new one rather than completely disappearing, which can be better seen in the animation anim\_combined\_small.mp4. The IEF channels show only little radial motion during their existence.

Figure \ref{fig4} shows the temporal evolution of the maximal unsigned velocity within the radial boundaries given by the black lines in Figure \ref{fig2} at nine azimuth angles. The LOS velocities of H$\alpha$ and \ion{Ca}{ii} IR closely match each other. Again, increased velocities indicating an IEF channel persist over usually at least 10\,min with some continuing over up to 60 min. We note that lateral motions of the IEF channels can lead to the disappearance of the velocity signature at a fixed azimuth angle in that case. 

\begin{figure*}
\resizebox{14cm}{!}{\includegraphics{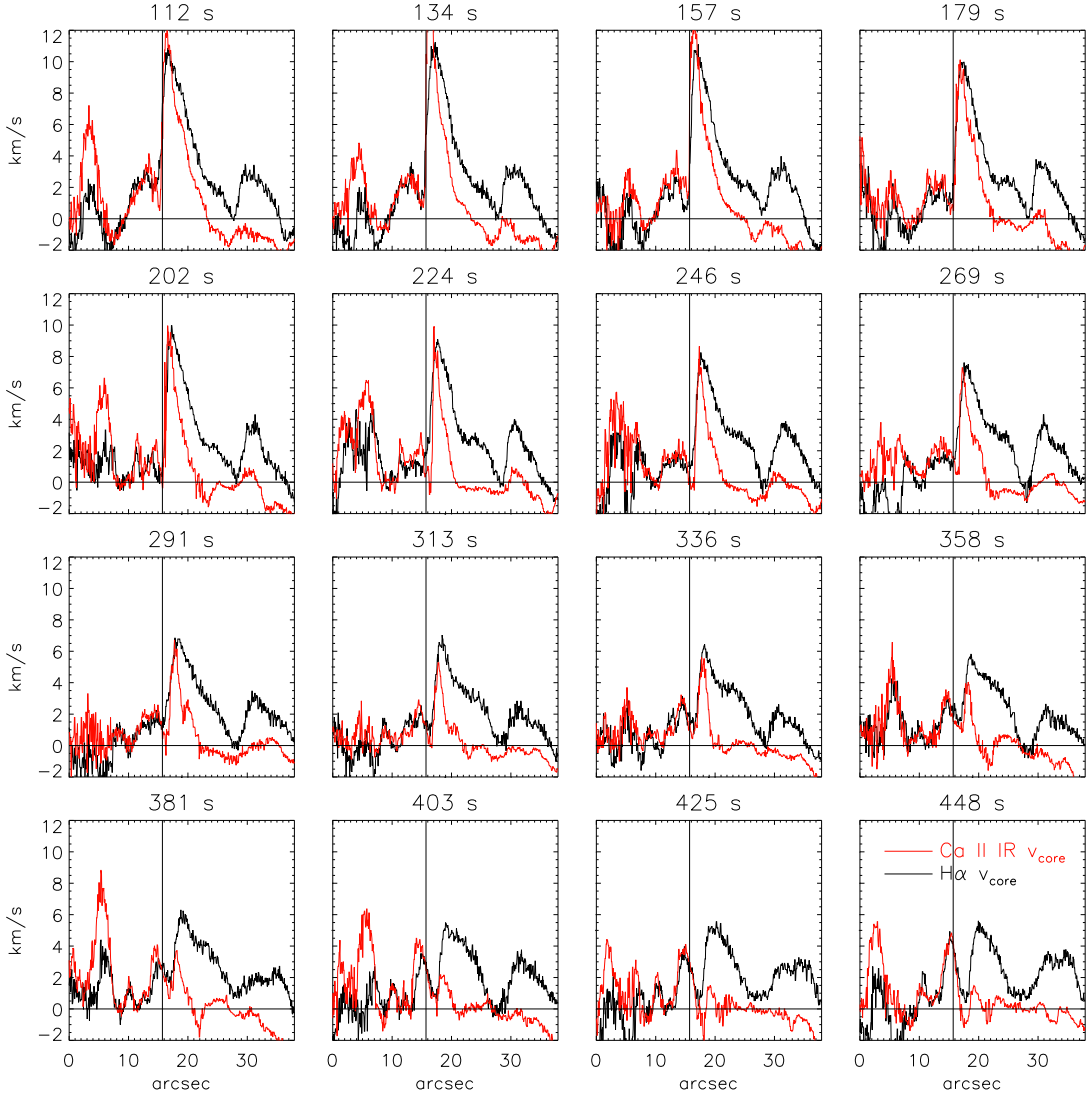}}\\$ $\\$ $\\
\caption{Temporal evolution of the LOS velocity at the end of an IEF channel at UT 14:44 at an azimuth angle of 351$^\circ$. Black/red lines: LOS velocity of H$\alpha$/\ion{Ca}{ii} IR. The black vertical lines mark the location of maximum flow speed at the first time step. Time increases from top to bottom across rows and left to right in each row.}\label{fig7}
\end{figure*}

As a third estimate of the life time of IEF channels we calculated the temporal auto-correlation of velocities across the time series. We restricted the range in azimuth to 0--90$^\circ$ and 270--360$^\circ$ for this calculation to avoid the presence of the long-lived filament structure on the limb side at an azimuth of about 200$^\circ$. We calculated the auto-correlation for three radial ranges covering the umbra, the locations of the IEF channels, and the sunspot moat beyond the upper radial boundary of the IEF locations. The temporal auto-correlation for the umbra shows a clear oscillatory pattern opposite to the IEF channels and the moat (left column of Figure \ref{fig5}). We thus averaged the auto-correlation over about one period of the umbral oscillations for a clearer picture (right column of Figure \ref{fig5}). In that form, it is obvious that the umbral auto-correlation drops fastest to about zero over less than 300\,s. The auto-correlation for the moat region decays slower and closely follows a purely exponential decay with a decay constant of 300\,s (e.g., the orange dotted line in the lower panel of Figure \ref{fig5}). The temporal auto-correlation for the region of the IEF channels decays faster than that of the moat region below 200\,s, but maintains an extended tail of higher correlation values from about 500\,s to 1800\,s, indicating again the persistence of some structures in the IEF channel region over time scales of 10\,min or more. \citet{maltby1975} determined a similar quantity as the auto-correlation that he defined as the number of IEF channels with unchanged velocity after a given time delay. We read off the corresponding values from his Figure 7, scaled them to be about unity at zero time delay, and over-plotted his results on the auto-correlation. The slope of his curve for $t < 400$\,s matches well with our result. Equalizing the values for long time delays $t>1500$\,s would make the match of the curves even closer.
\subsubsection{Start and End of IEF Channels}
With the constant evolution of the velocity pattern including the merging and splitting of IEF channels, it is rather difficult to unambiguously identify the start or end of a specific IEF channel. We manually identified four cases of the termination of halfway isolated IEF channels that did not get replaced instantly and two cases of the start of a flow. 

Figure \ref{fig6} shows the temporal evolution of one event of the start of an IEF channel that began at about UT 15:21 at an azimuth angle of 327$^\circ$. The  LOS velocity is not zero in the beginning from a previous IEF channel. The flow speed at the location marked with the vertical lines approximately doubles from about 6\,km\,s$^{-1}$ to 12\,km\,s$^{-1}$ and the increased velocity then persists for several minutes. There is almost no radial motion of the location of maximal velocity, with potentially a really small displacement towards the umbra. Figure \ref{fig7} shows the opposite case of a termination of an IEF channel at about UT 14:44 at an azimuth angle of 351$^\circ$. The flow speed reduces from about 12\,km\,s$^{-1}$ to 0\,km\,s$^{-1}$. The location of maximal velocity moves away from the umbra by a few arcseconds along with the decrease of the flow speed.
\begin{table}
\caption{Acceleration and deceleration of IEF speed.}\label{tab_speed}
\begin{tabular}{c|cccccc|c}\hline\hline
angle [deg] & 351 & 290 & 325 & 8 & 327 & 326 & --\cr
$a$ [m\,s$^{-2}$]& -80  &   -45   &  -45 & -36 &   18 & 14 & 45$^1$ \cr
0-8\,km\,s$^{-1}$ [s]& 100 & 179 &   179 & 224 & 448  & 560 & 179$^1$\cr
\end{tabular}\\$ $\\
1: Derived from \citet{montesinos+thomas1993aa}.
\end{table}
\begin{figure}
\hspace*{.8cm}\resizebox{8cm}{!}{\includegraphics{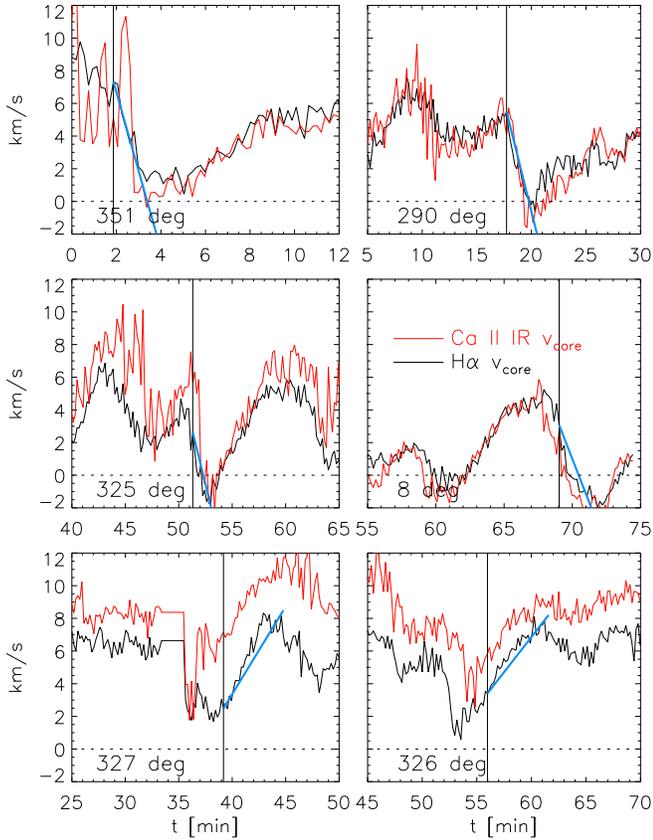}}\\$ $\\$ $\\
\caption{Temporal evolution of the LOS velocity at the start (bottom row) and end (top two rows) of IEF flow channels. Black/red lines: LOS velocity of H$\alpha$/\ion{Ca}{ii} IR. The black vertical lines mark the assumed time of the onset of the change of flow speed. The inclined blue lines indicate the acceleration and deceleration, respectively.}\label{fig8}
\end{figure}

Figure \ref{fig8} and Table \ref{tab_speed} summarize the temporal evolution of the six cases of start and end of an IEF channel that we manually identified. The deceleration at the end seems to be about twice as fast as the acceleration at the start of an IEF channel. The corresponding velocity would change from about sound speed to zero in less than 3\,min (bottom row of Table \ref{tab_speed}), while in the opposite direction it takes about 8--10 min to reach sound speed when starting from zero. The deceleration derived from the change in the LOS velocity at a fixed spatial location (Figure \ref{fig8}) is likely to be somewhat overestimated because of the additional outwards radial motion of the IEF downflow point, but in general the deceleration of an IEF channels seems to be faster than its acceleration. The acceleration falls short of the one corresponding to free fall by an order of magnitude. For comparison, we derived the corresponding acceleration for the steady-state solution of a critical siphon flow of \citet{montesinos+thomas1993aa} from the middle panel of their Figure 3. The flow in their case accelerates from about 3 to 6\,km\,s$^{1}$ over a distance of 300\,km with a linear velocity increase. Using the average velocity of 4.5\,km\,s$^{-1}$ for covering the distance yields an acceleration of about 45\,m\,s$^{-2}$, which matches our deceleration values.

\subsection{Locations of IEF channels}
In Paper I and \citet{beck+choudhary2019} we were only able to track 100 manually identified IEF channels in observations at a low cadence of about 20\,min. The current IBIS time series allows us to increase the statistics on the locations of IEF channels and downflow points to more than 10,000. 
\begin{figure}
\hspace*{.8cm}\resizebox{8.cm}{!}{\includegraphics{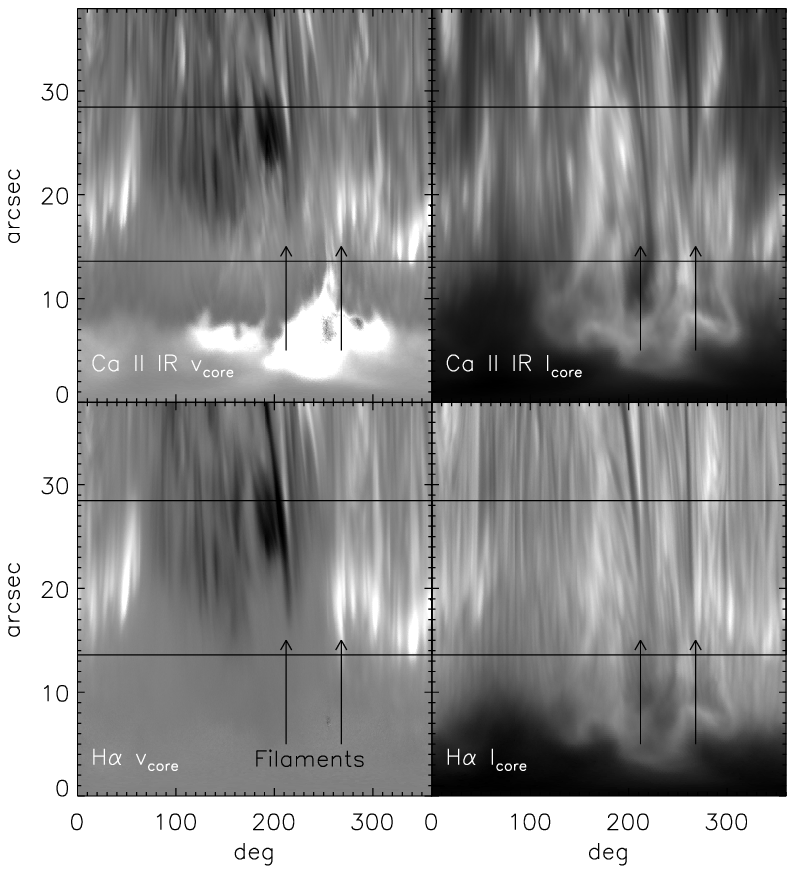}}\\$ $\\$ $\\$ $\\
\hspace*{.8cm}\resizebox{8.cm}{!}{\includegraphics{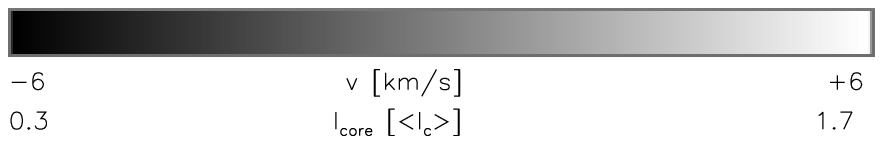}}

\caption{Temporal average over the time series. Bottom (top) row, left to right: line-core intensity and velocity of H$\alpha$ (\ion{Ca}{ii} IR). The horizontal black lines indicate the radial range used for the determination of the location of IEF channels. The black vertical arrows mark the locations of the two strongest superpenumbral filaments in the FOV.}\label{fig11}
\end{figure}
\begin{figure}
\resizebox{8.8cm}{!}{\includegraphics{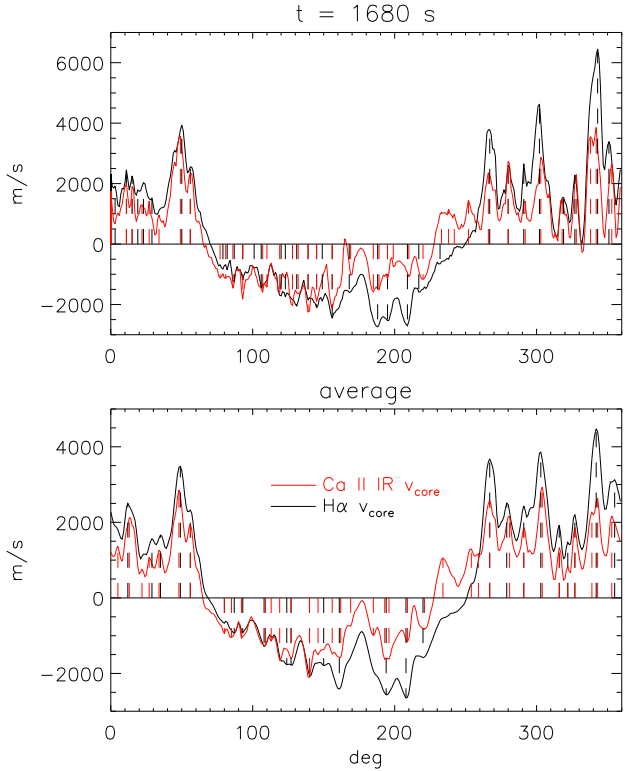}}\\
\caption{Determination of the location of IEF channels in azimuth angle. Bottom panel: determination of local velocity extrema in the temporal average. Black/red solid lines: temporal average of the velocity in H$\alpha$ and \ion{Ca}{ii} IR spatially averaged between the two black lines in Figure \ref{fig11}. Black/red vertical dashed lines: identified locations of enhanced flow speeds in H$\alpha$ and \ion{Ca}{ii} IR. Top panel: the same for an individual spectral scan with only the spatial averaging.}\label{fig12}
\end{figure}
\subsubsection{Azimuth Angles}
To determine the preferred azimuth locations of the IEF channels, we first averaged the radial cuts over the time series (see Figure \ref{fig11}). While the umbral oscillations average out completely both in the LOS velocity and the line-core intensity, several of the IEF channels that persist for a large fraction of the time still clearly stand out. There are no pronounced differences between H$\alpha$ and \ion{Ca}{ii} IR apart from the somewhat higher contrast in the \ion{Ca}{ii} IR line-core intensity. The same is valid for the comparison of the FIRS \ion{He}{i} data with the IBIS pseudo-scan maps in Figure \ref{fig1a}. 

\begin{figure}
\hspace*{.8cm}\resizebox{8.cm}{!}{\includegraphics{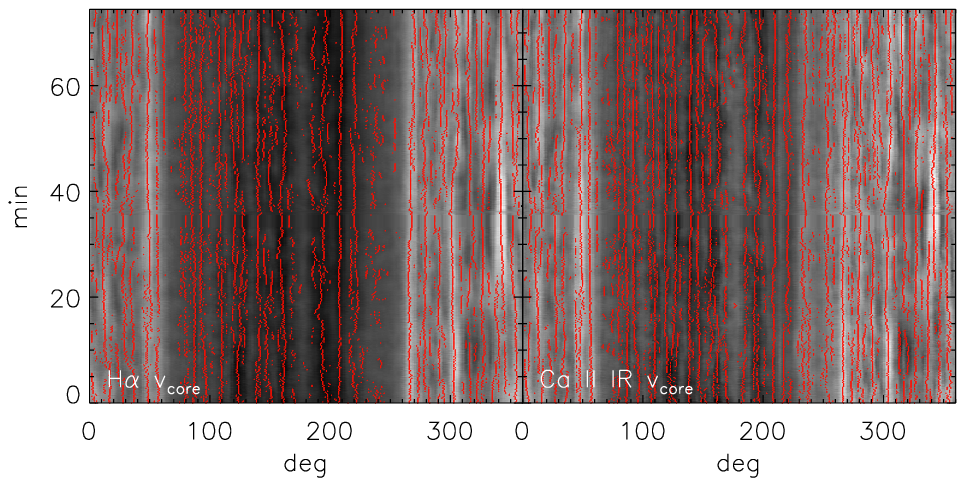}}\\$ $\\$ $\\$ $\\
\hspace*{.8cm}\resizebox{8.cm}{!}{\includegraphics{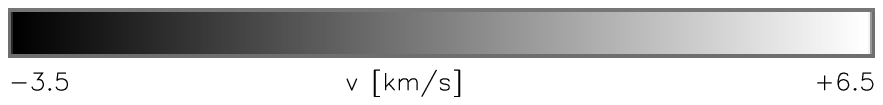}}
\caption{Locations of IEF channels with azimuth and time. Left/right panel: LOS velocity of H$\alpha$ and \ion{Ca}{ii} IR for each spectral scan averaged between the two black lines in Figure \ref{fig11}. Red dots: identified locations of local velocity extrema.}\label{fig13}
\end{figure}

We then averaged the velocities between the black horizontal lines in Figure \ref{fig11} in the radial direction and determined the locations of IEF channels in azimuth by identifying local extrema -- either minima or maxima -- in the azimuthal direction (bottom panel of Figure \ref{fig12}). The same approach was applied to the radial cuts of each individual spectral scan (top panel of Figure \ref{fig12}). On average, about 35 IEF channels were found in the azimuthal direction in each spectral scan, while there were 32 IEF channels in the temporally averaged velocity map. 

The locations of the IEF channels found in each spectral scan in Figure \ref{fig13} confirm again the presence of IEF channels at the same place over several 10 minutes up to the full duration of the time series. It can also be seen that the lateral motion in azimuth is small and usually does not exceed  2--3$^\circ$. The absence of IEF channels around 60$^\circ$ and 240$^\circ$ results from the lack of LOS velocities at those angles because of the projection effects.
\begin{figure}
\resizebox{8.8cm}{!}{\includegraphics{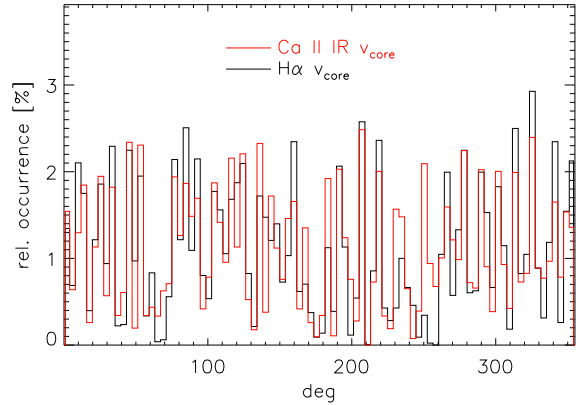}}
\caption{Histogram of the azimuth angles of IEF channels. Black/red line: derived from the velocity of H$\alpha$ and \ion{Ca}{ii} IR.}\label{fig14}
\end{figure}

Figure \ref{fig14} shows the histogram of the azimuth angles at which IEF channels were found in the full time series. There are several angles with an increased probability of hosting IEF channels that naturally coincide with the locations of high velocities in the average azimuthal velocity variation (Figure \ref{fig12}) or the time-dependent determination (Figure \ref{fig13}) that exhibit the existence of the long-lasting IEF channels. The spacing between such preferred azimuthal locations is about 10$^\circ$ in the histogram, which is fully compatible with the average number of about 35 IEF channels found in each spectral scan under the assumption of equidistant spacing.
\subsubsection{Relation to Moving Magnetic Features}
We only used a visual inspection of the aligned IBIS data and HMI magnetograms in Figure \ref{fig1} to investigate a potential relation of the IEF channels and their evolution to moving magnetic features (MMFs) in the moat around the sunspot, but there is no obvious connection between IEF channels and MMFs. On the one hand, the temporal scales of the evolution are completely off. The IEF channels change and evolve over less than 1 minute in some cases, while the MMFs in the FOV barely move by a few arcseconds over the full time series of one hour and also maintain their shape to the largest extent, e.g., compare the first and last panel of the bottom row of Figure \ref{fig1}. On the other hand, there are either almost no MMFs in the region with the most prominent IEF channels (black rectangle in Figure \ref{fig1}) at azimuth angles from 270--360$^\circ$ or the photospheric magnetic flux is comparably static (at azimuth angles from 0--90$^\circ$). Even thresholding the HMI magnetogram to lower magnetic flux levels did not reveal any type of significant changes in the photospheric magnetic field in that area. 
\begin{figure*}
\hspace*{.5cm}\resizebox{8.5cm}{!}{\includegraphics{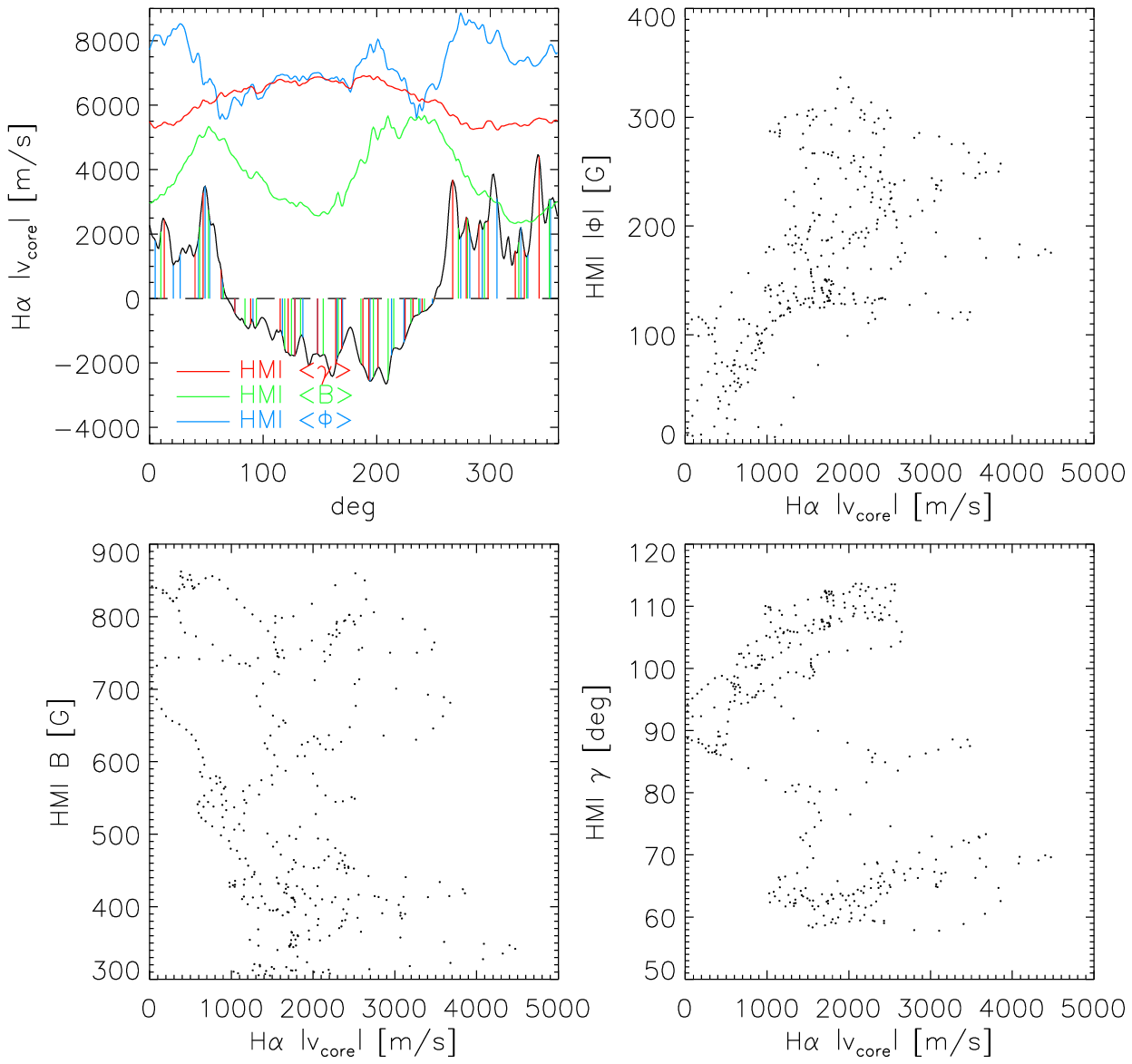}}\resizebox{8.5cm}{!}{\includegraphics{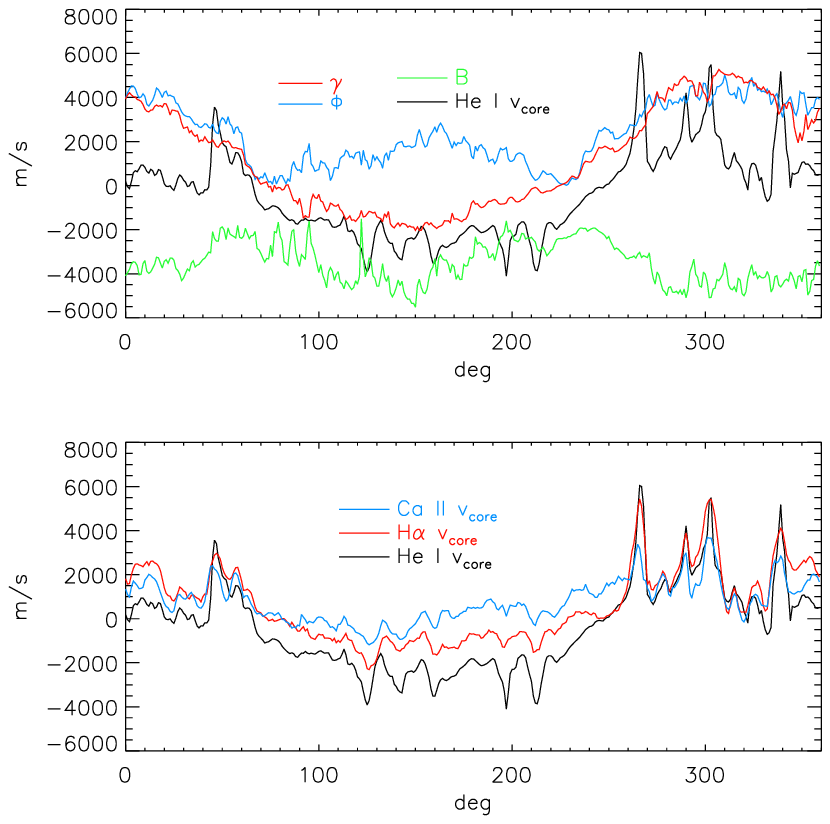}}\\$ $\\
\caption{Relations between IEF channel locations and penumbral magnetic field properties. Left four panels, top left: spatially and temporally averaged LOS velocity of H$\alpha$ (black line), HMI magnetic field inclination $\gamma$ (red line), HMI magnetic field strength B (green line) and HMI LOS magnetic flux $\Phi$ (blue line). The vertical red, green, and blue colored lines indicate the locations of local extrema in $\gamma$, B and $\Phi$. Left four panels, top right and bottom row: scatter plots of $\Phi$, B and $\gamma$ against the LOS velocity of H$\alpha$. Rightmost column, bottom panel: spatially averaged LOS velocities of \ion{He}{i} (black line), and from the pseudo-scan of H$\alpha$ (red line) and \ion{Ca}{ii} IR (blue line). Rightmost column, top panel: \ion{He}{i} LOS velocity (black line), magnetic field strength B (red line), LOS magnetic flux $\Phi$ (blue line) and inclination $\gamma$ (green line) from the SIR inversion of \ion{Si}{i} at 1082.7\,nm.}\label{fig15}
\end{figure*}
\subsubsection{Relation to Penumbral Magnetic Field}
To determine a possible relation between the preferred azimuthal locations of IEF channels and the penumbral magnetic field, we averaged the HMI magnetic field results first over the duration of our observations and then over some radial range inside the penumbra. We applied the same identification of local extrema in $B$, $\gamma$, and $\Phi$ as used on the LOS velocities for the identification of IEF channels. The upper left panel of Figure \ref{fig15} shows the resulting azimuthal variation of  $B$, $\gamma$, and $\Phi$ from HMI and the corresponding LOS velocity of H$\alpha$ with the identified locations of local extrema in the magnetic field properties overlaid. Even if there seems to be some overlap of locations with higher field strength or magnetic flux with the locations of IEF channels in the azimuthal curves, the connection is not convincing. The scatter plot of photospheric field strength against the chromospheric H$\alpha$ velocity shows no trend at all (lower left panel of Figure \ref{fig15}), while the correlation between LOS magnetic flux $\Phi$ and the LOS velocity of H$\alpha$ (top panel in second column) just results from the common property of passing through zero and becoming maximal at the same azimuthal angles. 

We repeated the same procedure with the FIRS \ion{Si}{i} inversion results and the IBIS pseudo-scan maps to check if the lack of correlation might be related to the spatial resolution or polarimetric sensitivity. We obtained, however, the same result with no obvious relation between photospheric magnetic field properties and the preferred locations of the IEF channels (top right panel of Figure \ref{fig15}), or in this case, even more clearer the actual absence of significant magnetic field variation at the preferred azimuthal locations of IEF channels. The characteristic azimuthal scale of variations in the photospheric penumbral field is also about 10$^\circ$, e.g., the green line in the top right panel of Figure \ref{fig15}, so any apparent co-spatiality of IEF channels and local magnetic field strength maxima in azimuth could well be fully coincidental. The majority of the IEF downflow points are located inside the penumbra (see next section).  
\subsubsection{Radial Distance from Spot Center}
\begin{figure}
\resizebox{8.8cm}{!}{\includegraphics{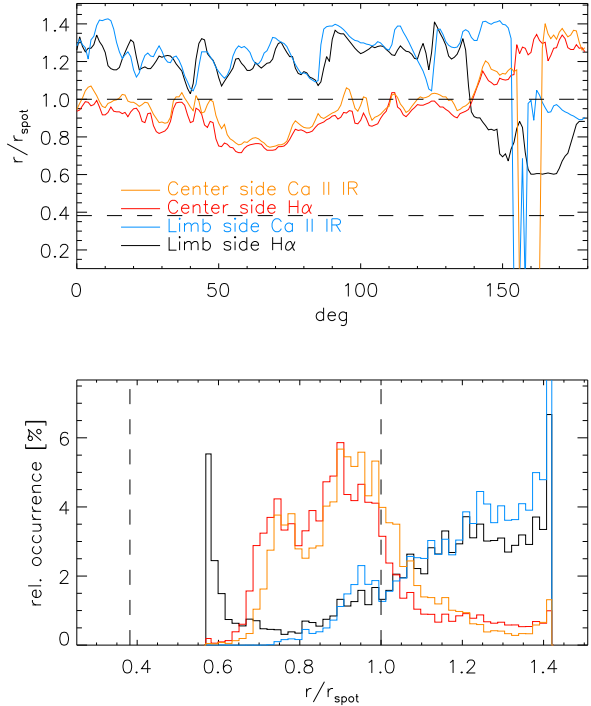}}
\caption{Radial distance of maximal flow velocities from the sunspot center. Top panel: temporal average of the radial distance of the location of maximal velocity between the two black lines in Figure \ref{fig11} in each spectral scan at each azimuth angle. Black/blue lines: derived from the LOS velocity of H$\alpha$ and \ion{Ca}{ii} IR on the limb side. Red/orange lines: derived from the LOS velocity of H$\alpha$ and \ion{Ca}{ii} IR on the center side. Bottom panel: histograms of the radial distance derived from the same quantities as above. The two black dashed lines indicate the umbral and penumbra boundary.}\label{fig16}
\end{figure}

For the determination of the radial distance of the IEF channels from the sunspot center we located the maximal velocity within the radial boundaries given in Figure \ref{fig11}. The maximal velocity within that radial range is usually found at the end of an IEF channel (Figures \ref{fig6} and \ref{fig7}). The bottom panel of Figure \ref{fig16} shows the histogram of the radial distances and the top panel the average value with azimuth angle. The limb and center side were treated separately to take the difference between them due to the projection effects into account. The IEF downflow points are found to lie from 0.6 to 1.4 spot radii on the center side, with a maximum probability of occurrence within the penumbra at around $0.9\pm0.14$ spot radii. For the limb side, the locations shift beyond the outer penumbral boundary and the probability monotonically increases with increasing radial distance. These values are, however, strongly biased by the location of the symmetry line of the sunspot that reduces the LOS velocities at the expected true location of the IEF downflow points to zero, and by the presence of the large-scale, curved, and permanently present filaments on the limb side of the sunspot (see Figure \ref{fig1}). 

\begin{figure}
\resizebox{8.8cm}{!}{\includegraphics{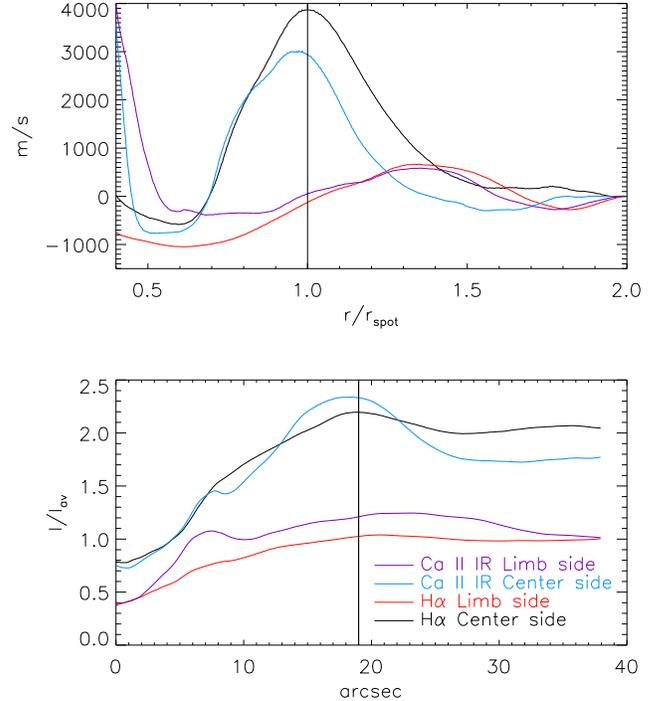}}
\caption{Average radial intensities (bottom row) and velocities (top row). Black and red lines: H$\alpha$ on center and limb side. Blue and purple lines: \ion{Ca}{II} IR on center and limb side. The vertical black lines indicate the outer penumbral boundary located at about 19$^{\prime\prime}$.}\label{temp_aver}
\end{figure}

As an independent determination of the radial location of the downflow points, we also directly averaged the LOS velocities and line-core intensities over time. Figure \ref{temp_aver} shows the radial variation of the H$\alpha$ and \ion{Ca}{ii} IR velocities and intensities separately for the limb and center side. The same difference between limb and center side as before is found with the maximal flow velocity further out on the limb side. The line-core intensities are less affected by the projection effects as there is no symmetry line for intensity. The downflow points with increased intensity are therefore found closer to the penumbra on the limb side as well. We note that the H$\alpha$ intensity also shows a radial intensity maximum at about the sunspot boundary and thus must be sensitive to temperature effects to some extent. 
\subsection{Resolved H$\alpha$, \ion{Ca}{ii} IR and \ion{He}{i} Spectra}
\begin{figure*}
\resizebox{6cm}{!}{\includegraphics{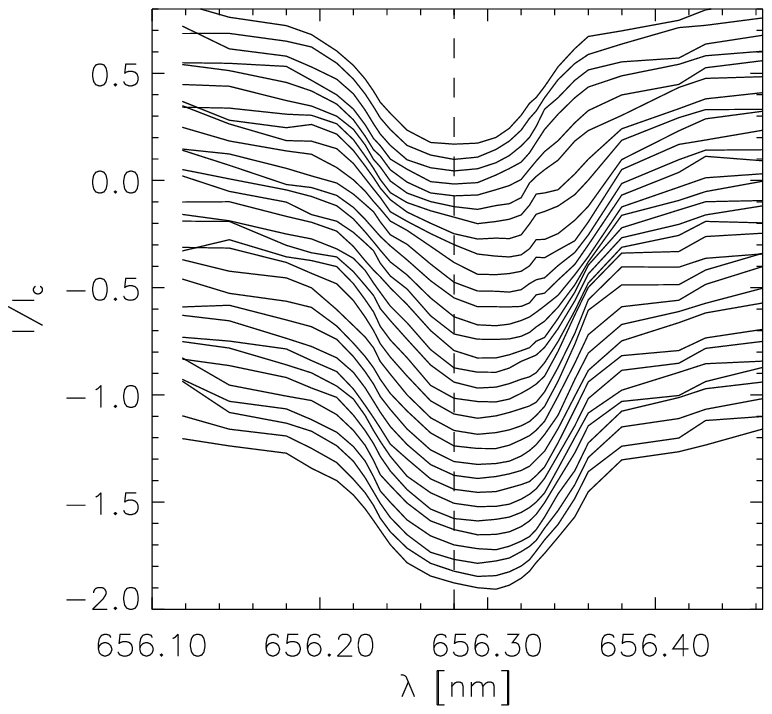}}\resizebox{6cm}{!}{\includegraphics{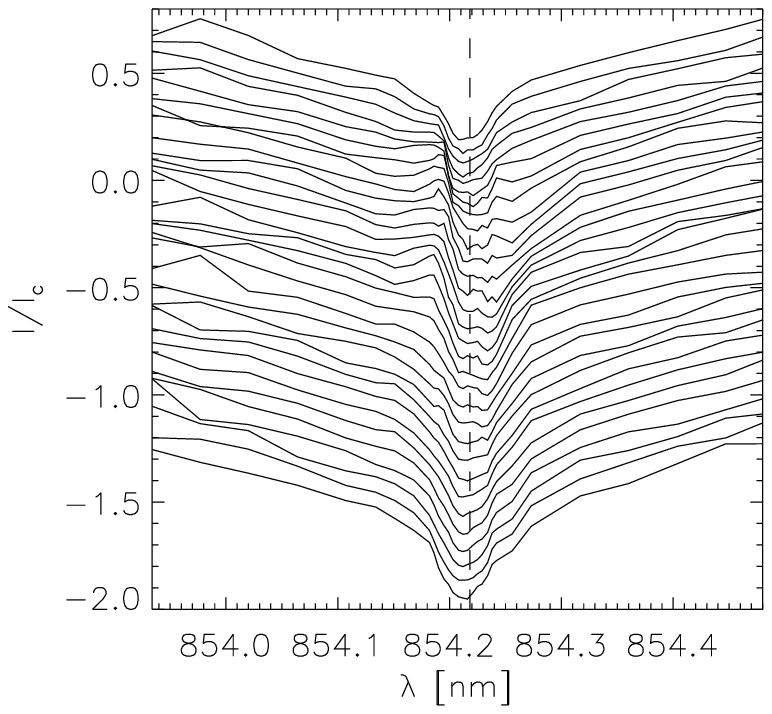}}\resizebox{6cm}{!}{\includegraphics{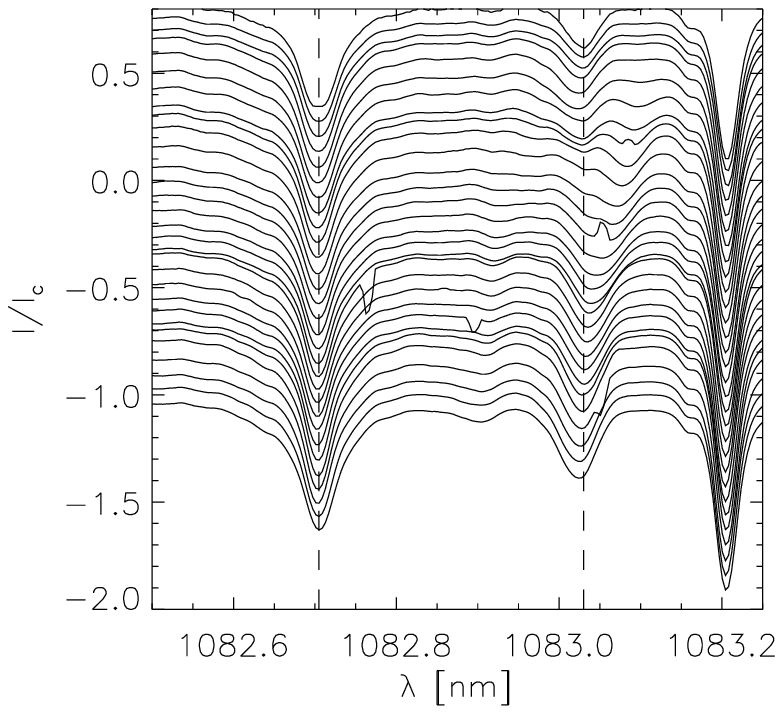}}
\caption{Example spectra of H$\alpha$ (left panel), \ion{Ca}{ii} IR (middle panel) and \ion{He}{i} at 1083\,nm (right panel) along an IEF channel around the downflow point. The dashed black vertical lines indicate the rest wavelength of the spectral lines. The downflow point is located in the upper part at a relative intensity of about zero and the umbra is towards the top of the panels.}\label{figspec}
\end{figure*}

In Paper I, we found that the IEF channels do only show up as a strongly Doppler-shifted line satellite close to the downflow points. Tests with synthetic data suggested that this results even for an assumed flow channel with 100\,\% fill factor in the upper atmosphere, which would indicate a vertical structuring along the LOS. To cross-check whether it still possibly could be caused by the limited spatial resolution of $> 0\farcs7$ of the SPINOR data used in Paper I, we extracted a similar set of individual spectra along an IEF channel from the current data sets. Figure \ref{figspec} shows H$\alpha$, \ion{Ca}{ii} IR and \ion{He}{i} spectra on a cut along a flow fibril that intersected the downflow point. For the IBIS data in H$\alpha$ and \ion{Ca}{ii} IR, we obtain the same general picture as before with a strong component at rest and a weak line satellite in the red line wing with large LOS velocities. The spectra of the \ion{He}{i} line at 1083\,nm differ to some extent, but upstream of the downflow point. For \ion{He}{i} the unshifted component is missing or weaker than the Doppler-shifted one there. Beyond the downflow point the pattern changes to the same as in the other lines, with the Doppler-shifted component getting weaker the larger the Doppler shift and the closer to the umbra the spectrum was recorded. The pattern in all lines therefore still would comply with a vertical structuring, where for \ion{He}{i} there is no absorbing matter of high opacity apart from the IEF channel radially outwards from the downflow point.

\section{Summary}\label{sec_summ}
We investigated a time series of H$\alpha$ and \ion{Ca}{ii} IR spectra of about one hour duration and a spatial map in the \ion{He}{i} line at 1083\,nm to infer the temporal evolution of the inverse Evershed flow (IEF). All three chromospheric lines show a very similar behavior in their line-core intensity and the LOS velocity with matching spatial and temporal properties. We find that individual IEF channels persist for a few ten minutes to more than one hour. IEF channels that disappear are often rapidly replaced by a new channel at about the same location after a short time. The IEF channels show little radial or lateral motion and usually end in the mid to outer penumbra. Initiation of the flow takes about 10\,min, while the termination is faster and takes only about 5\,min. The IEF channels seem to appear at preferred azimuth angles that are spaced at about 10$^\circ$ distance. We cannot find any clear relation of the locations of the IEF channels to local variations or specific properties of the photospheric penumbral magnetic field nor any causal connection of the IEF channels to moving magnetic features outside of the sunspot. Several more transient processes that can be seen in the time series were not studied in full detail, but in general seem to have only little impact on the structure and evolution of the IEF channels.
\section{Discussion}\label{sec_disc}
Our current results are in line with the findings of Paper I and of most previous literature. 
\subsection{Structure of the IEF} 
We find no significant difference between the appearance of the IEF channels in the three chromospheric lines of \ion{He}{i} at 1083\,nm, \ion{Ca}{ii} IR at 854\,nm and H$\alpha$ at 656\,nm. All spectra and the derived quantities comply with a picture of IEF channels as flow fibrils at chromospheric levels that turn down to the photosphere at the inner end point in the penumbra, where a shock forms because of the supersonic flow speed. Even at the improved spatial resolution as compared with Paper I, the IEF channels show up as a Doppler-shifted line satellite close to the downflow points. This suggest a vertical structuring along the LOS instead of lateral spatial resolution effects.

Averaged flow speeds of about 4\,km\,s$^{-1}$ (Figure \ref{temp_aver}) match the results of 4--6\,km\,s$^{-1}$ in for instance \citet{dialetis+etal1985} or \citet{georgakilas+etal2003}. Thanks to the higher spatial and spectral resolution of our data we can demonstrate that the true LOS velocities of individual IEF channels are 2--4 times as high, which makes them supersonic at photospheric heights already without additionally taking the projection effects onto the LOS into account. The radial locations of the downflow points inside the penumbra or close to the outer penumbral boundary also match previous findings \citep{dialetis+etal1985,alissandrakis+etal1988,dere+etal1990,georgakilas+etal2003}. We were not able to find any clear connection of the IEF channels in general or downflow points in specific to the local photospheric magnetic field properties or its temporal evolution neither in the penumbra nor in the moat, even if the IEF channels seem to occur repeatedly at preferred azimuthal angles around the sunspot center. 

The pronounced difference between LOS velocities on the center and limb side of the sunspot will be primarily due to the projection effects (left half of Figure \ref{figdiscus}) at the heliocentric angle of the observations of 43$^\circ$. Assuming field-aligned flows, the LOS is about perpendicular to the flow direction just where the IEF downflow points on the limb side would be. This is confirmed by the presence of an enhanced line-core intensity on the limb side at about the same radial distance as on the center side because the intensity is barely impacted by the LOS effects.
\subsection{Temporal Evolution of the IEF}
Our findings on the temporal evolution of the IEF channels closely follow those of \citet{maltby1975} and \citet{georgakilas+christopoulou2003}. The life time of individual IEF channels is of a few to a few ten minutes, while some last for the full duration of the observations of more than one hour. Similar to the description of \citet{maltby1975} we see occurrences of IEF channels getting replaced by new ones at about the same location. An established IEF channel persists with about the same properties for several minutes with little radial or lateral motion and without any clear impact of the transient events discussed in the next section. The IEF channels are clearly not oscillatory phenomena. In the few cases, where we could track the start and stop of individual isolated IEF channels, we find that the initial acceleration over up to about 10\,min takes about twice as long as the deceleration at the end. Accelerations are on the order of a few ten m\,s$^{-2}$ and fall short of gravitational acceleration in free fall. 
\subsection{Relation of IEF and Transient Events} 
Our observations had a higher spatial, spectral and temporal resolution than most previous data used for similar studies on the IEF. This allowed us a much clearer distinction between the IEF channels and other more transient phenomena that can clearly be seen to be present in the animation of the observation and multiple plots, but which were not investigated in detail in the current study.
\subsubsection{IEF Channels and Running Penumbral Waves}
From a preliminary analysis of running penumbral wave \citep[RPW; see, e.g.,][and citations and references therein]{giovanelli1972,zirin+stein1972,brisken+zirin1997,bloomfield+etal2007,jess+etal2013,priya+etal2018} events in our data, we found that their maximal LOS velocity amplitude is only about 1\,km\,s$^{-1}$ in and near the umbra. The amplitude of the RPWs decays rapidly with radial distance (lower row of Figure \ref{fig2}). At the location of the IEF downflow points the IEF velocity is usually about an order of magnitude larger than that of the RPWs. In the animation of the observations, especially in the temporal and azimuthal derivative of the H$\alpha$ LOS velocity, it can be seen that the RPWs that initially arrive as a spatially extended wavefront at the umbral boundary tend to split into multiple threads, some of which run along existing IEF channels. The IEF channels, however, persist without any visible change throughout multiple such events. We thus do not see any direct influence of RPWs on the IEF channels, also not in connection with the start and stop of an IEF channel. 

\begin{figure*}
\centerline{\resizebox{7.cm}{!}{\includegraphics{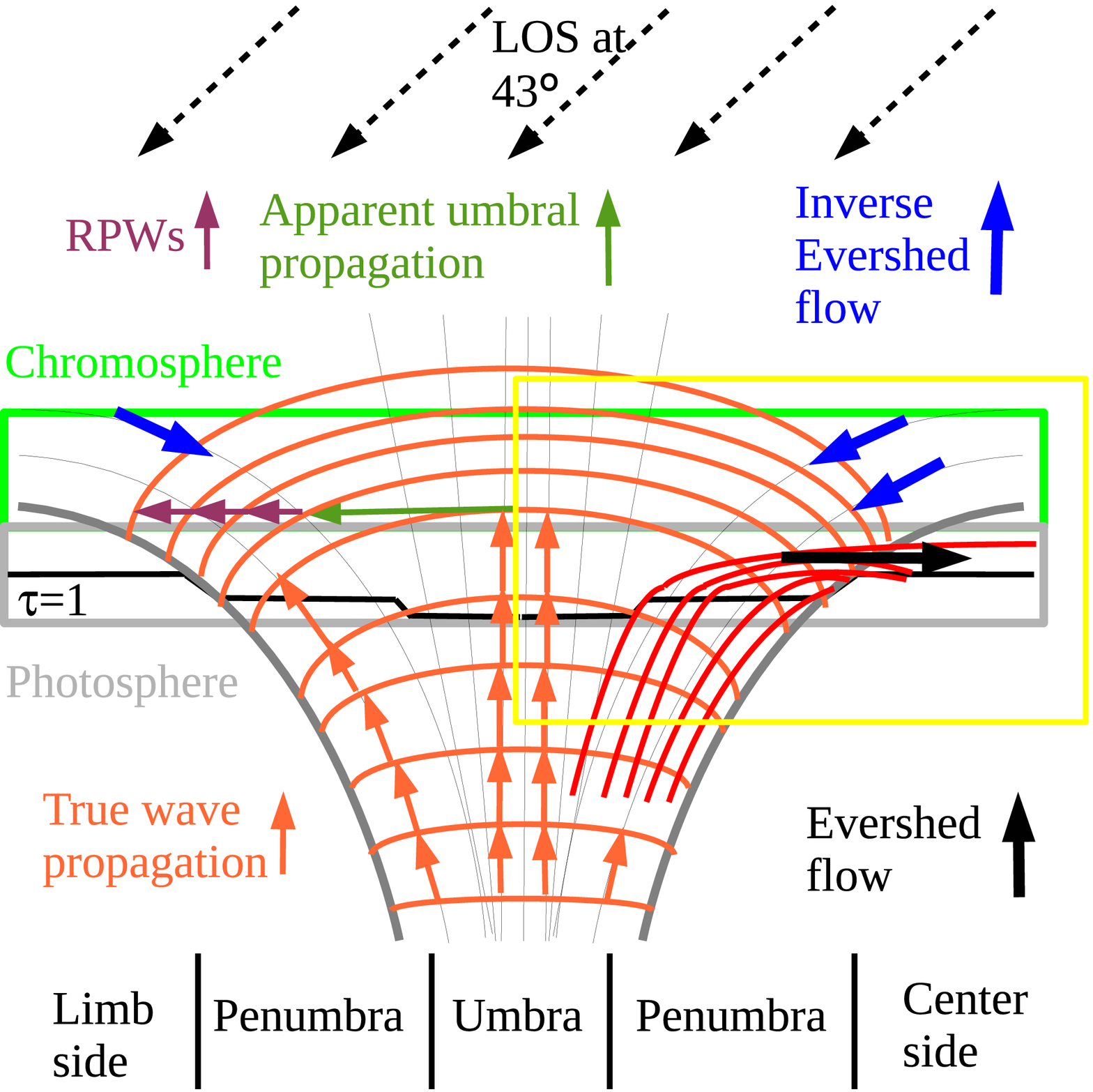}}\hspace*{1cm}\resizebox{7.cm}{!}{\includegraphics{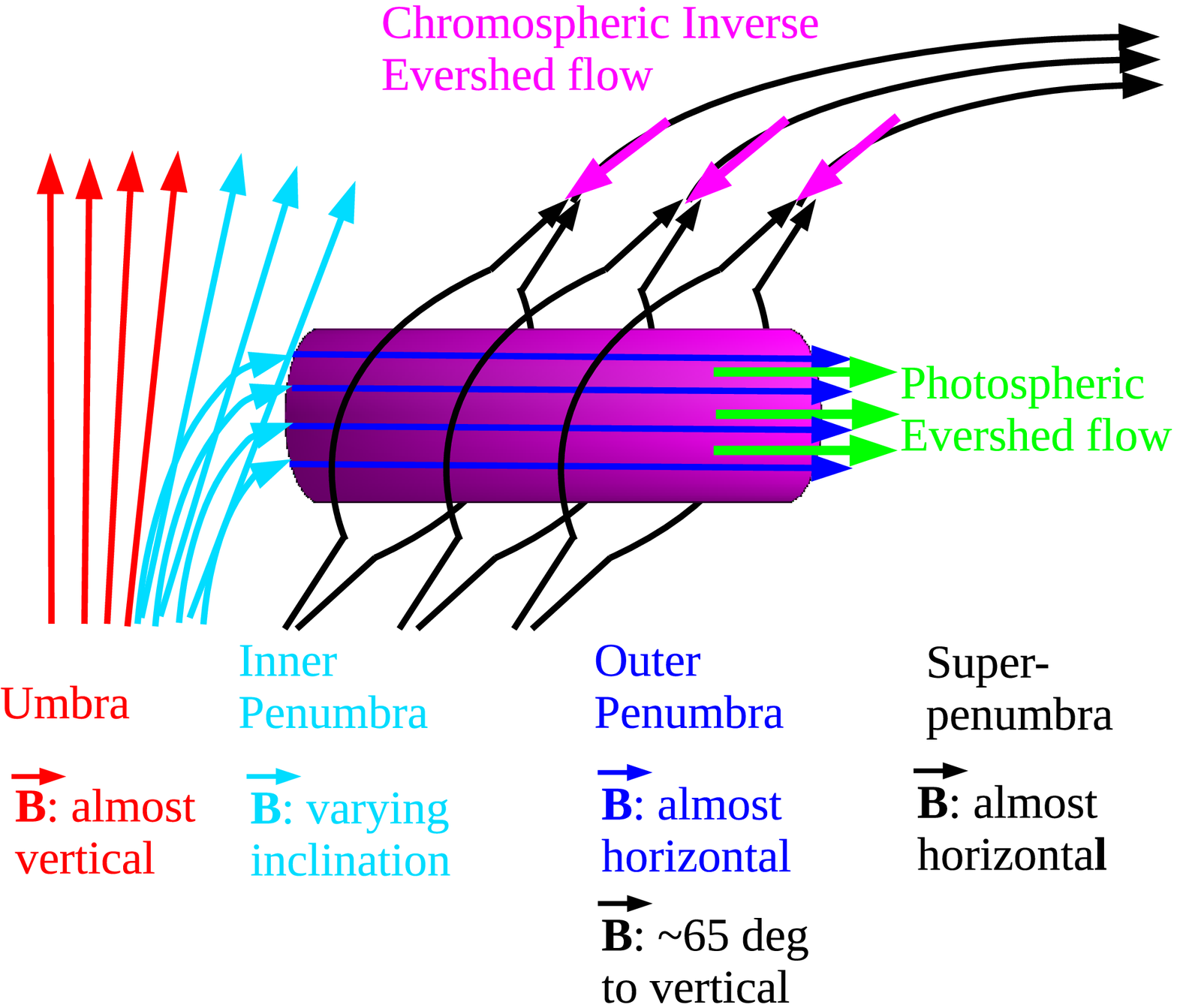}}}
\caption{Sketch of the magnetic topology of a sunspot and the various wave and flow phenomena. Left panel: side view of a sunspot. The grey and green rectangles indicate photosphere and chromosphere. Grey and red lines indicate magnetic field lines. Orange lines and arrows show the wavefront and assumed propagation of global sunspot oscillations. Black and blue arrows highlight the direction of the EF and IEF. The apparent radial propagation of umbral and penumbral waves is indicated by green and purple arrows. Right panel: magnification of the assumed magnetic field vector and the flows in the penumbra inside the yellow rectangle of the left panel.}\label{figdiscus}
\end{figure*}
\subsubsection{IEF Channels and Velocity Packets}
A second feature that we did not investigate in more detail are the events labeled ``velocity packets'' by \citet{georgakilas+christopoulou2003}. They consist of two types, small roundish patches of a LOS velocity with an opposite sign to the IEF located between the IEF downflow points and the umbra and elongated fibril-like patches radially outward from the downflow points (seen as black color inside the black rectangle in the LOS velocity maps of Figure \ref{fig1} and the animation). The roundish patches sometimes propagate radially outwards through the IEF downflow points and convert to the second type of elongated features. The latter seem to always move away from the sunspot.

Again, a preliminary analysis confirmed generally velocity amplitudes much smaller than those of the IEF channels. Similar to their reaction to RPWs, IEF channels also persist without any clear change when such velocity packets pass through or along them. Even looking at individual cases did not make it obvious what the actual physical process is that leads to the velocity packets. The elongated type outwards of the IEF downflow points gives the impression of rising loops or field lines in some cases. The evolution of the velocity packets is generally much faster than that of the IEF channels.

Given the lack of a clear relation to the photospheric magnetic field structure and evolution in and close to the sunspot we suggest that the primary driver governing the temporal evolution of IEF channels could be located at the outer end of the IEF channels in the sunspot moat instead. A reconfiguration of the magnetic field at the outer end could generate the conditions for the initialization or termination of the flow along specific IEF channels. The velocity packets could actually be related to such a process when they reach the outer end of the moat flow, but this requires further study.
\subsection{Relation to Global Sunspot Structure} 
Figure \ref{figdiscus} attempts to put our results on the IEF structure and evolution from \citet{choudhary+beck2018}, \citet{beck+choudhary2019} and the current study into the larger picture of sunspot topology and dynamics. 

The results on the inclined flow angle \citep{haugen1969,dialetis+etal1985,georgakilas+etal2003,beck+choudhary2019} and the direction opposite to the photospheric EF require that the IEF connects to a different set of magnetic field lines than those that harbor the nearly horizontal EF (right panel of Figure \ref{figdiscus}). Those field lines would have to reach chromospheric heights in or close to the penumbra, continue throughout the superpenumbra in or close to the chromosphere -- because the IEF channels can be identified over several Mms in the radial direction outside of the penumbra -- and then connect to places presumably at the end of the moat cell. Moving magnetic features would correspond to photospheric structures below the IEF channels that would connect to the horizontal magnetic field lines of the EF instead \citep{cabrerasolana+etal2006}. Umbral field lines would only reach the photosphere again at much larger distances from the sunspot and with much larger apex heights, which could render it impossible to drive a siphon flow along them. This topology would imply that EF and IEF are two distinct, unrelated phenomena with eventually different physical drivers. 

As far as the radial propagation of umbral waves (UWs) and RPWs is concerned, our data suggest that they most likely result from coherent oscillations across the whole sunspot cross-section below the photosphere. Each RPW in Figure \ref{fig2} has an associated UW \citep[see also][]{tziotziou+etal2006}. The apparent radial propagation speed of the UWs is 100\,km\,s$^{-1}$ or more, with in some case zero slope of the UW phase ridges in the $r-t$-diagram, which would imply infinite speed. A wave source below the surface across the whole sunspot cross-section would be able to consistently explain both the UW and RPW speeds, where the latter show an only apparent radial propagation caused by an increased time delay when reaching the chromosphere because of the longer distance to be traveled along the inclined penumbral magnetic field lines \citep[orange arrows in the left panel of Figure \ref{figdiscus};][]{bloomfield+etal2007}.

\subsection{Future Work}
Apart from the topics on the transient phenomena that merit further investigation, we would like to point out some more avenues that could be pursued on the IEF itself. 

The detailed relation between the EF and IEF, or its absence, requires simultaneous information on photospheric and chromospheric layers. The different bisector levels in our analysis, which we did not actually make use of in the current study, would allow to trace EF, IEF and wave phenomena with height in the atmosphere, but only within limits. A combination of quasi-simultaneous photospheric and chromospheric observations would be better suited. Of especial interest would be the search for any possible relation between the IEF downflow points and photospheric downflows in the penumbra \citep{katsukawa+etal2009,katsukawa+etal2010}. The IEF downflow points are well localized and the flow speed is not zero after the deceleration by the stationary shock (Paper I), so there is mass that should reach the photosphere, but only at a few well-defined places. An in-depth analysis of the height variation of thermal and dynamic properties would also benefit from more sophisticated approaches in the analysis of the spectra by means of an inversion of the \ion{Ca}{ii} IR and H$\alpha$ spectra \citep{beck+etal2019a,schwartz+etal2019}.

Another topic of interest would be to track the evolution at the outer end of IEF channels in or at the end of the moat cell. We were unable to identify trigger events that control the start or end of the IEF channels at the inner end point within the limits of the spatial resolution of our polarimetric data. Such data would require a FOV that covers both ends of the IEF channels in one go at a spatial resolution similar to that of the IBIS data used here, but including spectropolarimetry.
\section{Conclusions}\label{sec_concl}
We find that the inverse Evershed flow (IEF) happens along long-lived (10--60\,min) superpenumbral fibrils that show little evolution over their life time. The IEF is no oscillatory phenomenon, but a stationary flow along arched magnetic field lines that connect the outer penumbra with locations in or at the end of the sunspot moat. We did not find any relation of individual IEF channels to specific local photospheric magnetic field properties in the penumbra or to moving magnetic features outside of the sunspot. Fully in line with our previous results we propose that the IEF is driven by a siphon flow mechanism. At the spatial resolution of our polarimetric data, we cannot definitely determine if the events that govern the start or termination of IEF channels do happen at the outer end of the flow channels in the moat or at the inner end in the penumbra.
\begin{acknowledgements}
The Dunn Solar Telescope at Sacramento Peak/NM was operated by the National Solar Observatory (NSO). NSO is operated by the Association of Universities for Research in Astronomy (AURA), Inc.~under cooperative agreement with the National Science Foundation (NSF). IBIS has been designed and constructed by the INAF/Osservatorio Astrofisico di Arcetri with contributions from the Universit{\`a} di Firenze, the Universit{\`a} di Roma Tor Vergata, and upgraded with further contributions from NSO and Queens University Belfast. HMI data are courtesy of NASA/SDO and the HMI science team. This work was supported through NSF grant AGS-1413686. We thank S. Gosain (NSO) and S. Criscuoli (NSO) for helpful comments on the paper.
\end{acknowledgements}

\bibliographystyle{aa}
\bibliography{references_luis_mod}

\end{document}